\documentclass[12pt]{elsarticle}

\usepackage{mathstyle00}
\usepackage{amsmath}
\usepackage{ulem}
\DeclareMathOperator{\Tr}{Tr}
\makeatletter
\def\ps@pprintTitle{%
  \let\@oddhead\@empty
  \let\@evenhead\@empty
  \let\@oddfoot\@empty
  \let\@evenfoot\@oddfoot
}
\makeatother
\usepackage[margin=1.0in]{geometry}    
\usepackage{caption}
\usepackage{tikz}
\usetikzlibrary{matrix,shapes,arrows,positioning}
\usepackage{comment}

\newcommand{\R}{\mathbb{R}}

\newcommand{\eq}[1]{\begin{equation} #1 \end{equation}}
\newcommand{\eqs}[1]{\begin{align} #1 \end{align}}

\begin{document}
\begin{frontmatter}
		\title{A variational neural network approach for glacier modelling with nonlinear rheology}

		\author[mu]{Tiangang Cui}
		\ead{tiangang.cui@monash.edu}
        \author[uoc]{Zhongjian Wang\corref{cor1}}
        \ead{zhongjian@statistics.uchicago.edu}
		\author[hku]{Zhiwen Zhang\corref{cor1}}
		\ead{zhangzw@hku.hk}
		
		\address[mu]{School of Mathematics, Monash University, Victoria, Australia.}
		\address[uoc]{Department of Statistics and CCAM, The University of Chicago, Chicago, IL 60637, USA.}
		\address[hku]{Department of Mathematics, The University of Hong Kong, Pokfulam Road, Hong Kong SAR, China.}
		\cortext[cor1]{Corresponding author}
		
\begin{abstract}
We propose a mesh-free method to solve the full Stokes equation for modeling the glacier dynamics with nonlinear rheology. Inspired by the Deep-Ritz method proposed in [1], we first formulate the solution to the non-Newtonian Stokes equation as the minimizer of a variational problem with boundary constraints. Then, we approximate its solution space by a deep neural network. The loss function for training the neural network is a relaxed version of the variational form, in which penalty terms are used to present soft constraints due to mixed boundary conditions. Instead of introducing mesh grids or basis functions to evaluate the loss function, our method only requires uniform sampling from the physical domain and boundaries. Furthermore, we introduce a re-normalization technique in the neural network to address the significant variation in the scaling of real-world problems. Finally, we illustrate the performance of our method by several numerical experiments, including a 2D model with the analytical solution, the Arolla glacier model with realistic scaling and a 3D model with periodic boundary conditions. Numerical results show that our proposed method is efficient in solving the non-Newtonian mechanics arising from glacier modeling with nonlinear rheology.

\noindent \textit{\textbf{AMS subject classification:}} 35A15, 65J15, 68T99, 70K25, 76A05.   
 

\end{abstract}
		
\begin{keyword}
Deep learning method; variational problems; mesh-free method; non-Newtonian mechanics; nonlinear rheology;  glacier modelling.	
\end{keyword}
\end{frontmatter}

\section{Introduction}
\noindent
In recent years, deep neural networks (DNNs) have achieved unprecedented levels of success in a broad range of areas such as computer vision, speech recognition, natural language processing, and health sciences, producing results comparable or superior to human experts \cite{lecun2015deep,goodfellow2016deep}. The impacts have reached physical sciences where traditional first-principle based modeling and computational methodologies have been the norm. Thanks in part to the user-friendly open-source computing platforms from industry (e.g. TensorFlow and PyTorch), there have been vibrant activities in applying deep learning tools for scientific computing, such as approximating multivariate functions, solving ordinary/partial differential equations (ODEs/PDEs) and inverse problems using DNNs; see, e.g. \cite{Kutz_19,weinan2018deep,JinchaoXu:2018,khoo2017solving,Brenner_18,ZabarasZhu:2018,wang2020mesh}
 and references therein. 
There are many classical works on the approximation power of neural networks; see e.g. \cite{cybenko1989approximation,hornik1989multilayer,ellacott1994aspects,pinkus1999approximation}.  For recent works on the expressive (approximation) power of DNNs; see, e.g. \cite{cohen2016expressive,schwab2017deep,yarotsky2017error,QiangDU:2018,Shen_21a,Shen_21b}. In \cite{JinchaoXu:2018}, the authors  showed that DNNs with rectified linear unit (ReLU) activation function and enough width/depth contain the continuous piece-wise linear  finite element space. Thus, one can represent a solution of PDE using the ReLU-DNN.

Solving ODEs or PDEs by a neural network (NN) approximation is known in the literature dating back at least to the 1990's; see e.g. \cite{LeeH1990, Fernandez1994, Lagaris1998}. The main idea in these works is to train NNs to approximate the solution by minimizing the residual of the ODEs or PDEs, along with the associated initial and boundary conditions. These early works estimate neural network solutions on a fixed mesh. Recently DNN methods are developed for Poisson and eigenvalue problems with a variational principle characterization (deep Ritz,  \cite{weinan2018deep}), for a class of high-dimensional parabolic PDEs with stochastic representations \cite{han2018solving}, for advancing finite element methods \cite{Xu_mgnet,Cai_20,Cai_21}, for nonconvex energy minimization in simulating martensitic phase transitions \cite{chen2022deep}, and for learning and generating invariant measures of stochastic dynamical systems with parameters \cite{wang2022deepparticle}. The physics-informed neural network (PINN) method \cite{raissi2019physics} and a deep Galerkin method (DGM) \cite{sirignano2018dgm} compute PDE solutions based on their physical properties. For parametric PDEs, a deep operator network (DeepONet) learns operators accurately and efficiently from a relatively small dataset based on the universal approximation theorem of operators \cite{lu2019deeponet};
a Fourier neural operator method \cite{li2020fourier}  directly learns the mapping from functional parametric dependence to the solutions of a family of PDEs. 
In \cite{BaoZhou_20a,BaoZhou_20b}, weak 
adversarial network methods are studied for weak solutions and inverse problems, see also related studies on PDE recovery from data via DNN \cite{Dongb_18,Dongb_19,Xiu_19, Xiu_20} among others. 
In the context of surrogate modeling and uncertainty quantification (UQ), DNN methods include Bayesian deep convolutional encoder-decoder networks \cite{ZabarasZhu:2018}, deep multi-scale model learning \cite{Yalchin2018deep}, physics-constrained deep learning method \cite{zhu2019physics}, see also \cite{khoo2017solving,schwab2017deep,karumuri2019simulator,Karn2021} and references therein.

In this work, we present a deep learning method for solving problems in non-Newtonian mechanics that obey certain variational principles. In particular, we focus on nonlinear Stokes problems in which the viscosity  nonlinearly depends on the strain rate. This type of problems plays a fundamental role in modelling geodynamic processes, for instance, the dynamics of glaciers \cite{hutter2017theoretical,petra2012inexact} and mantle convection \cite{schubert2001mantle,mckenzie1984generation}. 
The solutions of these problems typically face a combination of challenges, such as the presence of local features  emerging from the nonlinear rheology, saddle point problems due to the incompressibility condition, complex problem domain with high aspect ratios, and high contrast boundary conditions. 
There has been a lot of effort in developing efficient numerical methods to address these challenges. 
For example, the work of \cite{brown2010efficient,elman2014finite,heuveline2007inf,isaac2015solution,schwab1999mixed,stenberg1996mixed} designed and analyzed efficient and accurate high-order finite element discretization schemes. Delicated adaptation strategies, see \cite{isaac2015solution,rudi2015extreme} and references therein, are developed to ensure that these high-order methods can successfully resolve local features of the flow field and a wide range of length scales. The resulting discretized nonlinear systems have to be solved using either Picard fixed-point iterations or Newton's method. Most of the recent high-performance solvers adopt modified Newton's iterations together with high-performance iterative linear solvers, see, e.g.,  \cite{brown2013achieving,fraters2019efficient,isaac2015solution,may2015scalable}, to obtain a superlinear convergence rate.



Our work is inspired by the deep Ritz method proposed in \cite{weinan2018deep} and our recent progress in developing deep learning method to solve interface problem \cite{wang2020mesh}. We first formulate the PDEs into a variational problem, which leads to the objective function used by the neural network training. The second part consists of penalty terms arising from the boundary constraint of the governing PDE. In real world models, e.g. glacier sliding, the scale of the problem domain and the scale of physical parameters vary significantly. To address these challenges, we introduce a normalizing layer following the input of neural networks and design a strategy to balance penalty factors due to different boundary conditions. Using this combination of strategies, our proposed deep learning method is capable to solve nonlinear Stokes problems with different geometric scales, different parameter scales and boundary conditions using a universal configuration of network parameters.

The rest of the paper is organized as follows. In Section 2, we introduce the background 
of the Non-Newtonian ice flow model. In Section 3, we review the basic idea of deep neural network and formulate the variational neural network approach for solving the 
Non-Newtonian ice flow model. In addition, we also discuss some details of the implementation of our method. In Section 4, we present numerical results to demonstrate the accuracy of our method. Section 5 offers some concluding remarks.

\section{Background}

\newcommand{\vel}{{\bf u}}
\newcommand{\stress}{{\boldsymbol \sigma}}
\newcommand{\stressd}{{\boldsymbol \tau_\vel}}
\newcommand{\grav}{{\bf g}}
\newcommand{\strain}{\dot{\boldsymbol \epsilon}}
\newcommand{\bI}{{\bf I}}
\newcommand{\bT}{{\bf T}}
\newcommand{\bn}{{\bf n}}
\newcommand{\bdel}{{\boldsymbol \delta}}

Since glaciers form one of the natural low-pass filters of atmospheric variability, modelling the mechanics of glaciers is instrumental in revealing slow changes in the climate system that might otherwise be obscured by short-term noise. We consider a canonical glacier model that treats the flow of ice as non-Newtonian, viscous, incompressible, and isothermal fluid in the steady-state \cite{dukowicz2010consistent}. In this section, we first present the strong form of the ice flow model, and then discuss its variational formulation that naturally yields the weak form of the ice flow model. 

\subsection{Non-Newtonian ice flow model}
For an open, bounded domain $\Omega \subset \R^d$, we denote the velocity field (measured in meters per calendar year, i.e., $\mathrm{m} \, \mathrm{a}^{-1}$) and the stress tensor (measured in $\mathrm{Pa}$) of ice flow by $\vel=(u_1,...,u_d)^\top: \Omega \mapsto \R^d$ and $\stress_\vel: \Omega \mapsto \R^{d \times d}$ , respectively. The conservation laws of momentum and mass state that
\begin{align}
- \nabla \cdot \stress_\vel & = \rho \grav, \label{eq:momentum1}\\
\nabla \cdot \vel & = 0, 
\end{align}
where $\rho$ is the ice density ($910 \,\mathrm{kg} \,\mathrm{m}^{-3}$) and $\grav$ is gravitational acceleration ($9.81 \,\mathrm{m} \,\mathrm{s}^{-2}$, where $\mathrm{s}$ represents a second). The stress tensor is split into a deviatoric part and an isotropic pressure $p$, i.e., \(
\stress_\vel = \stressd - p \bI.
\)
Denoting the strain rate tensor of the velocity field by
\begin{align}
\strain_\vel = \frac12( \nabla \vel + \nabla \vel^\top),
\end{align}
the Glen's law of rheology \cite{paterson1994physics} links the deviatoric stress to the strain rate via the constitutive equation
\begin{align}
\stressd = 2 \eta(\vel)  \strain_\vel \quad \text{with} \quad  \eta(\vel) = \frac12 A^{-\frac{1}{n}} \left( \frac12 \strain_\vel : \strain_\vel \right)^{\frac{1-n}{2n}},\label{eq:nonlinear_rheology} 
\end{align}
where $\eta(\vel)$ is the effective viscosity nonlinearly depending on the strain rate and the double-dot product is the Frobenius inner product defined as
\[
  \boldsymbol \sigma_1 : \boldsymbol \sigma_2 = {\rm trace}(\boldsymbol \sigma_1^\top \boldsymbol \sigma_2^{}) = \sum_{i,j} (\boldsymbol \sigma_1)_{ij} (\boldsymbol \sigma_2)_{ij},
\]
for two second order tensors $\boldsymbol \sigma_1$ and $\boldsymbol \sigma_2$. Here $A = 10^{-16} (\mathrm{Pa}^{-n} \mathrm{a}^{-1}$) is the flow parameter 
and we often set $n=3$.  
Using the nonlinear constitutive equation \eqref{eq:nonlinear_rheology}, the conservation of momentum \eqref{eq:momentum1} becomes
\eq{
- \nabla \cdot \stressd + \nabla p = \rho \grav . \label{eq:momentum2}
}

Without loss of generality, we assume the domain $\Omega$ is bounded by two disjoint surfaces, $\Gamma_\mathrm{b}$ the bottom boundary and $\Gamma_\mathrm{t} = \partial \Omega \setminus \Gamma_\mathrm{b}$ the top boundary. Let $\bn$ denote the unit outward normal vector at any point on the boundary $\partial \Omega$. 
On the top boundary we impose the traction-free boundary condition
\begin{align}
  \stressd \bn - p_0 \bn = 0   \quad \mathrm{on}\quad \Gamma_\mathrm{t},
\end{align}
where $p_0$ is the atmospheric pressure. Since  the atmospheric pressure is often assumed to be negligible \cite{pattyn2008benchmark}, so that we adopt $p_0 = 0$ here.  
On the bottom boundary we impose a no-penetration condition along the outward normal direction, i.e., 
\begin{align}
  \vel \cdot \bn = 0 \quad \mathrm{on}\quad \Gamma_\mathrm{b},
\end{align}
and a sliding boundary condition along the tangential direction. Given the map to the tangential direction $\bT = \bI - \bn \otimes \bn$, the sliding boundary condition is given as
\begin{align*}
  \bT \big( \stressd \bn - p \bn \big) + \beta \bT \vel = 0 \quad \mathrm{on}\quad \Gamma_\mathrm{b},
\end{align*}
for some basal drag coefficient $\beta > 0$. Note that the basal drag coefficient $\beta$ is a function of location in general. 
Since 
\(
\bT \bn = 0 , 
\)
the sliding boundary condition does not depend on the pressure, which can be reduced to
\begin{align}
  \bT \big( \stressd \bn \big) + \beta \bT \vel = 0 \quad \mathrm{on}\quad \Gamma_\mathrm{b}.
\end{align}
Along the tangential direction, there is a special case that $\beta \rightarrow \infty$, where the sliding boundary condition becomes a non-sliding Dirichlet boundary condition, i.e., 
\(
\bT \vel = 0.
\) In later numerical examples, our algorithm can be adapted to both cases.

To summarize, the ice flow with nonlinear rheology can be modelled by a system of equations
\begin{align}
- \nabla \cdot \stressd + \nabla p & = {\bf f} \quad \mathrm{on} \quad  \Omega\label{eq:strong1}\\
\nabla \cdot \vel & = 0 \;\;\quad  \mathrm{on} \quad \Omega\label{eq:strong2}\\
\stressd \bn - p_0 \bn & = 0 \;\;\quad   \mathrm{on} \quad \Gamma_\mathrm{t} \label{eq:strong3}\\
\vel \cdot \bn & = 0 \;\;\quad  \mathrm{on} \quad \Gamma_\mathrm{b} \label{eq:strong4}\\
\bT \big( \stressd \bn \big) + \beta \bT \vel & = 0 \;\;\quad  \mathrm{on} \quad  \Gamma_\mathrm{b}, \label{eq:strong5}
\end{align}
where $\beta > 0$, ${\bf f} = \rho \grav$, $\bT = \bI - \bn \otimes \bn$, and $\stressd$ given in \eqref{eq:nonlinear_rheology}. With $\beta \rightarrow \infty$, the mixed boundary conditions in \eqref{eq:strong4} and \eqref{eq:strong5} simply become $\vel = 0$ on $\Gamma_\mathrm{b}$.

\subsection{Variational formulation}

The momentum equation \eqref{eq:strong1} and the divergence-free mass equation \eqref{eq:strong2} have the variational form: finding a vector function $\vel\in H^1(\Omega)$ satisfying boundary conditions \eqref{eq:strong3} and \eqref{eq:strong4} and a scalar function $p \in L^2(\Omega)$ satisfying $p = p_0$ on $\Gamma_\mathrm{t}$ such that
\eqs{
\int_\Omega \big( - \nabla \cdot \stressd + \nabla p - {\bf f} \big) \cdot {\bf v}  dV & = 0 \label{eq:weak1_mom} \\ 
- \int_\Omega q\, \nabla \cdot \vel dV & = 0  , \label{eq:weak1_mass}
}
for all vector functions ${\bf v}\in H^1(\Omega)$ satisfying ${\bf v} \cdot \bn = 0$ on $\Gamma_\mathrm{b}$ and ${\boldsymbol \tau}_{\bf v} {\bf n} - p_0 {\bf n} = 0$ on $\Gamma_\mathrm{t}$ and all scalar function $q \in L^2(\Omega)$ satisfying $q = p_0$ on $\Gamma_\mathrm{t}$. 
Applying the divergence theorem for vectors and tensors, the momentum equation \eqref{eq:weak1_mom} can also be expressed as
\eq{
\int_\Omega \left(\stressd : \strain_{\bf v} - p \nabla \cdot {\bf v} - {\bf f} \cdot {\bf v}\right) dV - \oint_{\partial \Omega} \big(\stressd \bn - p \bn\big) \cdot {\bf v} dS = 0.\label{eq:weak2}
}
The solutions $(\vel, p)$ satisfying the above variational form can be expressed as the minimizer of some energy functional subject to the divergence-free constraint \eqref{eq:weak1_mass} and some boundary conditions \eqref{eq:strong3}-\eqref{eq:strong5}. This provides a starting point for applying deep neural network approximation to model the glacier flow. We present the energy functional and the corresponding optimization problem for the sliding and the the non-sliding boundary conditions as follows. 


\paragraph{Sliding bottom boundary} 
Subject to the traction free boundary condition \eqref{eq:strong3} on the top boundary, the boundary integral at the top boundary is eliminated, i.e.,  
\eq{
 \oint_{\Gamma_{\rm t}} \big(\stressd \bn - p \bn\big) \cdot {\bf v} dS = \oint_{\Gamma_{\rm t}} \big(\stressd \bn - p_0 \bn \big)\cdot {\bf v} \, dS = 0.\label{eq:weak3}
}
where $p_0$ is the atmospheric pressure at the top boundary. 
At the bottom boundary, applying the condition ${\bf v} \cdot \bn = 0$ on $\Gamma_\mathrm{b}$ and the sliding boundary condition \eqref{eq:strong5} we have
\eqs{
  \oint_{\Gamma_{\rm b}} \big(\stressd \bn - p \bn\big) \cdot {\bf v} dS & = \oint_{\Gamma_{\rm b}} \big(\stressd \bn\big)\cdot {\bf v}  \, dS \nonumber \\
  & = \oint_{\Gamma_{\rm b}} \Big( \bT \big(\stressd \bn\big)\Big)\cdot \Big( \bT {\bf v} \Big) \, dS + \oint_{\Gamma_{\rm b}} \Big( (\bI - \bT) \big(\stressd \bn\big)\Big)\cdot \Big( (\bI - \bT) {\bf v} \Big) \, dS \nonumber \\
  & = - \oint_{\Gamma_{\rm b}} \beta \big(\bT \vel\big) \cdot \big(\bT {\bf v}\big) \, dS ,\label{eq:weak4}
}
as $\mathrm{range}(\bT)$ and $\mathrm{range}(\bI - \bT)$ are mutually orthogonal and $(\bI - \bT) {\bf v} = 0$.
Therefore, the boundary integral in \eqref{eq:weak2} can be simplified as
\[
- \oint_{\partial \Omega} \big(\stressd \bn - p \bn\big) \cdot {\bf v} dS  =  \oint_{\Gamma_{\rm b}} \beta \big(\bT \vel\big) \cdot \big(\bT {\bf v}\big) \, dS .
\]

As shown in \cite{dukowicz2010consistent}, the solutions $(\vel, p)$ satisfying the variational form \eqref{eq:weak2} can be considered as the minimizer of the the energy functional
\eq{ \mathcal{E}_\mathrm{s}(\vel) =  \int_{\Omega} \left( \frac{2n}{1+n}  A^{-\frac{1}{n}}  \left( \frac12 \strain_\vel : \strain_\vel \right)^{\frac{1+n}{2n}} - \rho \grav \cdot \vel \right) dV + \frac12 \oint_{\Gamma_{\rm b}} \beta \big(\bT \vel\big) \cdot \big(\bT \vel\big) \, dS,\label{eq:energy_stokes}}
subject to the divergence-free condition \eqref{eq:strong2}. Formulating this constrained optimization problem using the method of Lagrange multiplier, we have the Lagrangian functional 
\eq{  
\mathcal{L}_\mathrm{s}(\vel, p) =  \mathcal{E}_\mathrm{s}(\vel) - \int_{\Omega} p \nabla \cdot \vel dV,
}
in which the pressure function $p$ plays the role of the Lagrange multiplier. Since the directional derivative of $f(\vel) := \frac12\strain_\vel : \strain_\vel $ along a function $\bf v$ is $ f'(\vel)[{\bf v}] = \strain_\vel : \strain_{\bf v}$,
the variation of $\mathcal{L}(\vel,p)$ along functions $(\bf v,q)$ takes the form
\[
\mathcal{L}'_\mathrm{s}(\vel, p)[{\bf v}, q] = \int_\Omega \left(\stressd : \strain_{\bf v} -  p \nabla \cdot {\bf v} - \rho \grav \cdot {\bf v} \right) dV - \int_\Omega q\, \nabla \cdot \vel dV + \oint_{\Gamma_{\rm b}} \beta \big(\bT \vel\big) \cdot \big(\bT {\bf v}\big) \, dS.
\]
The solution to the Lagrange multiplier, $(\vel, p)$ such that $\mathcal{L}'_\mathrm{s}(\vel, p)[{\bf v}, q] = 0$ for all $({\bf v}, q)$ is equivalent to the solution to the variational form defined in  \eqref{eq:weak1_mom} and \eqref{eq:weak1_mass}. Thus, the velocity field $\vel$ of the ice flow model can be obtained by minimizing the energy functional $\mathcal{E}_\mathrm{s}(\vel)$ subject to the divergence-free constraint, the traction-free boundary condition \eqref{eq:strong3}, and the Dirichlet boundary conditions \eqref{eq:strong4} and \eqref{eq:strong5}.

\paragraph{Non-sliding bottom boundary} With non-sliding boundary condition at the bottom, i.e., $\vel = 0$ on $\Gamma_\mathrm{b}$, we do not need to impose the boundary condition \eqref{eq:strong5} using the boundary integral as the sliding boundary case in \eqref{eq:energy_stokes}. Following a similar derivation as above, the velocity field $\vel$ of the ice flow model with non-sliding boundary can be obtained by minimizing the energy functional 
\eq{ \mathcal{E}_\mathrm{ns}(\vel) =  \int_{\Omega} \left( \frac{2n}{1+n}  A^{-\frac{1}{n}}  \left( \frac12 \strain_\vel : \strain_\vel \right)^{\frac{1+n}{2n}} - \rho \grav \cdot \vel \right) dV, \label{eq:energy_stokes_ns}}
subject to the divergence-free constraint, the traction-free boundary condition \eqref{eq:strong3}, and the Dirichlet boundary conditions $\vel = 0$ on $\Gamma_\mathrm{b}$. The corresponding Lagrangian functional becomes
\eq{  
\mathcal{L}_\mathrm{ns}(\vel, p) =  \mathcal{E}_\mathrm{ns}(\vel) - \int_{\Omega} p \nabla \cdot \vel dV.
}

\subsection{Divergence-free solutions spaces} Instead of searching for the saddle point solution of the Lagrangian functionals, we can construct divergence-free solution spaces for the velocity field $\vel$ to reduce the constrained optimization problems to unconstrained optimization problems. In the two dimensional case, i.e., $(x,y) \in \Omega \subset \R^2$, we can define a potential function $\phi: \Omega \mapsto \R$, which leads to a vector function 
\[
\vel = \left[\frac{\partial \phi}{\partial y}, - \frac{\partial \phi}{\partial x}\right]^\top,
\]
that satisfies the divergence-free condition by construction. In the three dimensional case, i.e., $(x,y,z) \in \Omega \subset \R^3$, we can define a vector function $\boldsymbol\phi: \Omega \mapsto \R^3$, so that
\[
\vel = \nabla \times \boldsymbol\phi
\]
satisfies the divergence-free condition. Using the divergence-free construction of the velocity field, we can directly minimize the energy functional $\mathcal{E}_\mathrm{s}(\vel)$ and $\mathcal{E}_\mathrm{ns}(\vel)$ subject to appropriate boundary conditions without using the Lagrangian formulation. 

\section{Formulation of the DNN method}

\noindent
We first discuss the background of DNNs and then develop the DNN-based methods for solving the Non-Newtonian ice flow problems.

\subsection{DNNs and its approximation property}
\noindent
There are two ingredients in defining a DNN. The first one is a linear map of the form $T:R^n\rightarrow R^m$, defined as $T(x)=Ax+b$, where $A=(a_{ij})\in R^{m\times n}$, $x\in R^{n}$ and $b\in R^m$. The second one is a nonlinear activation function $\sigma:R\rightarrow R$. Common examples of the activation function include the rectified linear unit (ReLU), $\sigma(x)=\max(0,x)$, the soft-plus function, $\sigma(x) = \log(e^{x}+1)$, the sigmoid function, $\sigma(x)=(1+e^{-x})^{-1}$, etc. The definition of activation function can be trivially extended to a nonlinear map $\sigma: R^n\rightarrow R^n$ by applying the scalar-valued activation function element-wise to each of its inputs.

Using above definitions, we are able to define a continuous function $F(x)$ by a composition of linear transforms and activation maps, i.e., 
\begin{equation}\label{eqn:eg3layernet}
F(x)=T^{k}\circ\sigma\circ T^{k-1}\circ\sigma
\cdot\cdot\cdot\circ T^{1}\circ\sigma\circ T^{0}(x),
\end{equation}
where $T^{i}(x)=A_ix+b_i$ with $A_i$ and $b_i$ are unknown matrices and vectors to be estimated. Dimensions of $A_i$ and $b_i$ are chosen to make \eqref{eqn:eg3layernet} meaningful. Such a DNN is called a $(k+1)$-layer DNN, which has $k$ hidden layers. Collecting all the unknown coefficients (e.g., $A_i$ and $b_i$) in \eqref{eqn:eg3layernet} as $\theta\in\Theta$, where $\theta$ is a high-dimensional vector and $\Theta$ is the space of $\theta$. The DNN representation of a continuous function can be viewed as 
\begin{align}\label{eqn:solution_DNN}
F=F(x;\theta).
\end{align}
We use $\mathbb{F}=\{ F(\cdot,\theta)|\theta\in\Theta\}$ to denote the set of all expressible functions by the DNN parameterized by $\theta\in\Theta$. Then $\mathbb{F}$ provides an efficient way to represent unknown continuous functions, comparing with a linear solution space used in classic numerical methods, e.g., a trial space spaced by linear nodal basis functions in the FEM. In the sequel, we shall discuss the approximation property of the DNN, which is relevant to the study of the expressive power of a DNN model \cite{cohen2016expressive,schwab2017deep}. 

Early investigations of the approximation property of neural networks can be found in \cite{cybenko1989approximation,hornik1989multilayer}, where the authors analyzed the approximation property for the function class given by a feed-forward neural network with a single hidden layer. Later, many authors analyzed the error estimates of such neural networks in terms of the number of neurons, the number of network layers, and the choice of activation function; see \cite{ellacott1994aspects,pinkus1999approximation} for a throughout review of relevant works.

In recent years, DNNs have shown successful applications in a broad range of problems, including classification of complex systems and construction of response surfaces for high-dimensional models. Significant efforts have been devoted to investigate the dependence of the expressive power of DNNs on the network configuration. For example, in \cite{cohen2016expressive}, the authors proved that convolutional DNNs were able to express multivariate functions given in so-called hierarchical tensor formats. In \cite{yarotsky2017error}, the author studied the expressive power of shallow and deep neural networks with piece-wise linear activation functions and established new rigorous upper and lower bounds for the network complexity in approximating Sobolev spaces. In \cite{JinchaoXu:2018}, the authors proved that the continuous piece-wise bilinear finite element space is embedded in the space of DNNs constructed from the ReLU activation function, a sufficient width and a sufficient depth. Thus, we can uses DNNs to approximate the solution space spanned by the FEM basis. 



\subsection{Formulation of the variational neural network approach}\label{sec:deepRitz}
 
The model problem of the ice flow with nonlinear rheology \eqref{eq:strong5} can be solved by using numerical methods such as FEMs and FDMs. However, the nonlinearity in the problem bring essential difficulty in solving \eqref{eq:strong5} by the 
traditional numerical methods. Inspired by the 
recent development of deep learning based numerical method for solving variational problems \cite{weinan2018deep,wang2020mesh}, we will develop a variational DNN method to solve the glacier model problem defined in \eqref{eq:strong5}.

According to the divergence free property, i.e. $\nabla\cdot \vel=0$, we can represent the velocity field $\vel$ as follows:
\begin{align}\label{eq:helmoltz}
    \left[\frac{\partial \phi^0}{\partial y}, - \frac{\partial \phi^0}{\partial x}\right]^\top \text{ in 2D ~~~ and} \quad \nabla\times\begin{pmatrix}\phi^1\\\phi^2\\\phi^3\end{pmatrix} \text{ in 3D},
\end{align}
where $\phi^i$ is a scalar-valued function approximated by $F^i({\bf x,\theta})$. 
Here, $F^i({\bf x,\theta})$ is a DNN representation with ${\bf x}\in \R^d$ as its input and scalar output defined in \eqref{eqn:eg3layernet}. Moreover, $\theta$ denotes all the parameters that will be determined during the training stage.  Denoting the DNN representation of the velocity field by $\tilde\vel$, the numerical solution of \eqref{eq:strong5} can be obtained by finding $\theta\in\Theta$ that minimizes the energy functional $\mathcal{E}_\mathrm{s}(\tilde\vel)$ subject to various boundary conditions. 

Ideally, we want to define $\tilde\vel$ using $\theta\in\Theta_b \subset \Theta$, where  $\Theta_b$ is the maximal parameter subset such that the resulting $\tilde{\vel}$ satisfies the traction-free boundary condition \eqref{eq:strong3}, and the Dirichlet boundary conditions \eqref{eq:strong4} and sliding bottom boundary condition \eqref{eq:strong5}. 
After parameterizing the expressible function space by $\theta\in\Theta_b$, we equivalently define the 
variational problem \eqref{eq:energy_stokes} as
\begin{equation}\label{eqn:lag_representation_para}
 \min_{\theta\in\Theta_b}J(\theta)=\mathcal{E}_\mathrm{s}(\tilde{\vel}) =  \int_{\Omega} \left( \frac{2n}{1+n}  A^{-\frac{1}{n}}  \left( \frac12 \strain_{\tilde{\vel}} : \strain_{\tilde{\vel}} \right)^{\frac{1+n}{2n}} - \rho \grav \cdot \tilde{\vel} \right) dV + \frac12 \oint_{\Gamma_{\rm b}} \beta \big(\bT \tilde{\vel}\big) \cdot \big(\bT \tilde{\vel}\big) \, dS.
\end{equation}
Note that the divergence-free condition in \eqref{eq:weak1_mass} is automatically satisfied by the representation \eqref{eq:helmoltz}, and thus no additional treatment is needed.

The variational problems \eqref{eqn:lag_representation_para} is not convex in general and the integrals in \eqref{eqn:lag_representation_para} do not have a closed-form expression. Thus we numerically approximate the integrals by the Monte Carlo method and use the stochastic gradient descent (SGD) method \cite{bottou2010large} to minimize the objective function after Monte Carlo approximation. 
Denoting $\theta_k$ the $k$th component of the high-dimensional vector $\theta$, the derivative of $J(\theta)$ with respect to $\theta_k$ can be approximated as
\begin{align}
	\frac{\partial J\big(\theta\big)}{\partial \theta_k}
	\approx&\frac{\text{vol}(\Omega)}{N}\sum_{i=1}^{N}\partial_{\theta_k}\left( \frac{2n}{1+n}  A^{-\frac{1}{n}}  \left( \frac12 \strain_{\tilde{\vel}} : \strain_{\tilde{\vel}} \right)^{\frac{1+n}{2n}}(x_i)
	- \rho \grav \cdot \tilde{\vel} (x_i)\right)\nonumber \\
	&+\frac{\text{area}(\Gamma_b)}{N_b}\sum_{j=1}^{N_b}\partial_{\theta_k}\Big(\frac12 \beta \big(\bT \tilde{\vel}\big) \cdot \big(\bT \tilde{\vel}\big) (y_j) \, \Big)\label{num_lag},
\end{align} 
where random samples $x_i\overset{i.i.d.}{\sim} \text{Unif}(\Omega)$ are uniformed drawn from the physical domain $\Omega$, random samples $y_j\overset{i.i.d.}{\sim} \text{Unif}(\Gamma_b)$ are uniformed drawn from the bottom boundary $\Gamma_b$, $\text{vol}(\Omega)$ is the volume of the domain and $\text{area}(\Gamma_b)$ is the area of bottom boundary. In the context of deep learning method, $N$ and $N_b$ are called batch numbers, which mean the number of training examples utilized in one iteration. 

After approximating the gradient of $J(\theta)$, we can update each component of $\theta$ as follows: 
\begin{equation}\label{eqn:sgd_update}
	\theta_k^{n+1} =  \theta_k^{n} - \eta \frac{\partial J(\theta)}{\partial \theta_k}|_{\theta_k=\theta_k^{n}},
\end{equation} 
where $ \eta$ is the learning rate. To accelerate the training of the neural network, we use the Adam version of the SGD method \cite{kingma2014adam}.


In the objective function \eqref{eqn:lag_representation_para}, it is rather challenging to restrict the neural network parameter $\theta$ to the subset $\Theta_b$ that satisfies the boundary conditions, because the boundary of the subset may have complicated geometry. To address this issue, we adopt a relaxation approach by imposing boundary conditions as penalty terms. For each of the boundary conditions in \eqref{eq:strong3}-\eqref{eq:strong5}, we define a linear map $\mathcal{B}_j$ that maps the solution $\tilde u$ to the residual of the boundary constraint, which defines a boundary integral
\[
B_j(\theta) = \int_{{\partial D}_j}\big\| \mathcal{B}_j\tilde{u}(x,\theta) \big\|^2 dx
\]
for the boundary ${\partial D}_j$.
Then, the boundary conditions can be imposed as soft constraints via penalty terms to the objective functional $J(\cdot)$ in \eqref{eqn:lag_representation_para}. This leads to a new objective functional
\begin{equation}\label{eqn:lag_representation_bdd}
\tilde{u}_\varepsilon=\argmin_{\theta\in \Theta}\Big(J(\theta)+ \sum_{j = 1}^{b}\frac{1}{\varepsilon_j}B_j(\theta) \Big),
\end{equation} 
where $b$ is the number of boundary conditions needs to be imposed. 
Note that when a penalty term $\varepsilon_j^{-1}B_j(\theta)$ approaches zero as $\varepsilon_j \rightarrow 0$, the corresponding boundary condition is satisfied in the weak sense. 

\subsection{Implementation details} 
In addition to the conventional DNN formulation for solving PDE, we also made some modification that are shown to be essential in resolving the scale of the problem domain and the scale of physical parameters in real world problems.

\paragraph{Normalizing Layer} 
In some glacier model problems, periodic boundary conditions are used. For example, periodic boundary conditions are imposed in the two horizontal directions in the in the 2D model presented in Example C of the benchmarks \cite{pattyn2008benchmark}. To this end, we add an extra layer between input and first dense layer of our network. That maps $(x,y)$ to terms like $\big(\cos(2\pi x/T),\sin(2\pi x/T),\cos(2\pi y/T),\sin(2\pi y/T)\big)$,  where $T$ is the period in the horizontal dimensions. 

In most of the glacier modelling problems, the vertical scaling can be two or three orders of magnitude smaller than the horizontal scaling. To resolve this scaling, we also introduce a reparametrization: $x \mapsto x/L$, where $L$ is the diameter of the domain in corresponding dimension, to the normalization layer from input. 
Resolving this scaling is essential to ensure the expressibility of neural networks for the problems we are targeting here, because nonlinearity can be represented by a neural network may only exist in a bounded input domain. Taking a scalar single layer perception with the Sigmoid activation function  $\sigma(z) = (1 + e^{-z})^{-1}$ as an example, we have 
\[
f(x)=W_2 \sigma(W_1x+b_1)+b_2.
\] 
When $W_1x+b_1 $ is sufficiently large (e.g., $W_1x+b_1 \gg 5$),  $f(x)$ becomes a  constant approximately.  Thus, to sufficiently express nonlinear functions, scale of $W_1$ has to be $\mathcal{O}(L^{-1})$ without any reparametrization. However, most of existing implementations of machine learning toolboxes only support single-precision float-point numbers. This leads to inaccurate representation of $W_1$ and the gradient of the objective function, as the length scale of most of the real-world problems can be $O(10^4) - O(10^6)$. The reparametrization is critical to ensure the expressibility of neural networks and the numerical stability in these real-world problems.

\paragraph{Balancing penalty parameters of boundary conditions} Except for periodic boundary conditions, as shown in \eqref{eqn:lag_representation_bdd}, the other boundary conditions are enforced by penalty terms. However, there is no general framework to choose the penalty weight $\varepsilon_j$ in the context of DNN.  
It may not be suitable to use the same penalty parameter $\varepsilon_j$ for different boundary conditions, as the magnitude of the boundary maps $B_j(\theta)$ can vary significantly. 
For example, in the 3D model presented in Section 4.3, the term $B_j(\theta)$ is $O(10^{-4})$ for the traction-free condition at the top boundary and $O(10^{6})$ for the basal friction condition at the bottom boundary. Thus, finding the correct penalty parameter $\varepsilon_j$ for each of the boundary conditions is critical in balancing the soft constraints due to various boundary conditions. 
We propose to first estimate the scaling relations between each of the boundary integrals $B_j(\cdot)$ with the objective functional $J(\cdot)$ at the initialization stage of the network training. 
Denoting the initial parameter by $\theta_0$, we can then assign $\varepsilon_j$ as
\[
\varepsilon_j= \frac1{\varepsilon_0} \frac{B_j(\theta_0)}{ J(\theta_0)},
\]
so that the objective functional $J(\cdot)$ and each of the boundary integrals $B_j(\cdot)$ can be approximately balanced during the training. Here the common factor $\varepsilon_0$ is chosen to be $50$, which is an empirical constant shown to be sufficient in previous research \cite{jingrun_deepritz} and in our numerical experiments on simpler nonlinear models.

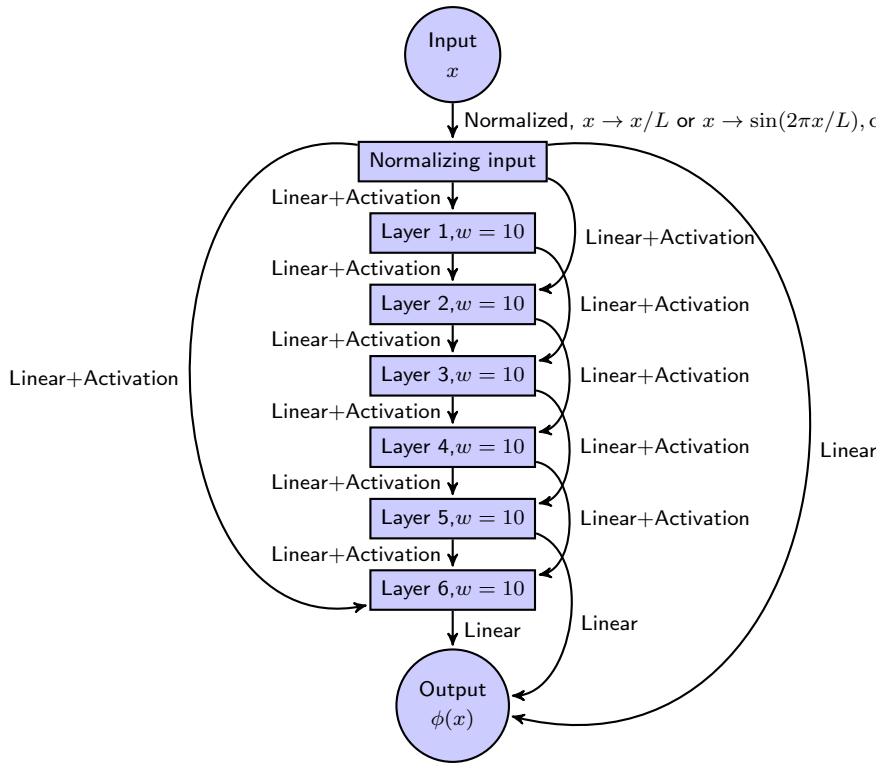
\begin{figure}[htbp]
	\begin{center}
		\begin{tikzpicture}[->,>=stealth',shorten >=1pt,auto,node distance=0.95cm, thick,io node/.style={circle,fill=blue!20,draw,  font=\sffamily\scriptsize,align=center},square node/.style={rectangle,fill=blue!20,draw,  font=\sffamily\scriptsize,align=center}]
		\node[io node] (x) {Input\\ $x$};
		\node[square node] (xnorm) [below = 0.5cm of x] {Normalizing input};
		\node[square node] (l1) [below of=xnorm] {Layer 1,$w=10$};
		\node[square node] (l2) [below of=l1] {Layer 2,$w=10$};
		\node[square node] (l3) [below of=l2] {Layer 3,$w=10$};
		\node[square node] (l4) [below of=l3] {Layer 4,$w=10$};
		\node[square node] (l5) [below of=l4] {Layer 5,$w=10$};

		\node[square node] (l6) [below of=l5] {Layer 6,$w=10$};

		\node[io node] (u) [below = 0.5cm of l6] {Output \\ $\phi(x)$};
		
		\path[every node/.style={font=\sffamily\scriptsize,fill=white,inner sep=1pt}]
		(x) 
		edge [] node[right=1mm] {Normalized, $x\to x/L$ or $x\to \sin(2\pi x/L),\cos(2\pi x/L)$} (xnorm)
		(xnorm) 
		edge [] node[left=1mm] {Linear+Activation} (l1)
		edge [bend left=80] node[right=1mm] {Linear+Activation} (l2)
		edge [bend right=100,looseness=1.3,overlay] node[left=1mm] {Linear+Activation} (l6)
		edge [bend left=100,looseness=1.7,overlay] node[right=1mm] {Linear} (u)
		(l1) 
		edge [] node[left=1mm] {Linear+Activation} (l2)
		edge [bend left=80] node[right=1mm] {Linear+Activation} (l3)
		(l2) 
		edge [] node[left=1mm] {Linear+Activation} (l3)
		edge [bend left=80] node[right=1mm] {Linear+Activation} (l4)
		(l3) 
		edge [] node[left=1mm] {Linear+Activation} (l4)
		edge [bend left=80] node[right=1mm] {Linear+Activation} (l5)
		(l4) 
		edge [] node[left=1mm] {Linear+Activation} (l5)
		edge [bend left=80] node[right=1mm] {Linear+Activation} (l6)
		(l5) 
		edge [] node[left=1mm] {Linear+Activation} (l6)
		edge [bend left=80] node[right=1mm] {Linear} (u)
		(l6) 
		edge [] node[right=1mm] {Linear} (u);
		\end{tikzpicture}			
	\end{center}
	\caption{The network Layout of $\phi$. }
	\label{DNN_phi}
\end{figure}

\paragraph{Network Hyperparameters}
In the numerical experiments of this work, we will apply the same network to represent $\phi$ in both two and three dimensions. After the normalizing layer, it has six latent dense layers with width 10. Between these dense layers we apply the Sigmoid activation function to guarantee the smoothness of our representation.  There is no activation function between the second last layer and the output layer. We illustrate the structure of our network in Fig.~\ref{DNN_phi}.

\section{Numerical Results}\label{sec:NumericalExamle}
In this section, we shall present numerical results in solving non-Newtonian Stokes equations to demonstrate the performance of the propose method. 
First we consider a two dimensional synthetic model with an analytical solution to demonstrate the efficiency and accuracy of our method. Then we apply the developed method on realistic benchmark problems \cite{pattyn2008benchmark} to demonstrate its generality. This includes a two dimensional model of the Arolla glacier and a three dimensional box model.

\subsection{A 2D Model on irregular domain}

We start with a two dimensional model defined on a domain that is enclosed by
\[
y = 0 \quad \text{and} \quad y = \frac12 x \left( 1 - x\right)\quad \text{with} \quad x \in [0, 1].
\]
To setup the benchmark, we start with a ground truth potential function
\[
\phi = \exp(x) (x-2)^2 y(y-1)^2,
\]
which leads to an analytical expression of the velocity field
\begin{align*}
    \left\{ \begin{array}{l}
         u=\exp(x)(x-2)^2(1-4y+3y^3)\\
         v=-\exp(x)(x-2)^2y(y-1)^2
    \end{array}\right.. 
\end{align*}
The above velocity field satisfies the conservation of mass condition in Eq.~\eqref{eq:strong2} and the no-penetration boundary condition in Eq.~\eqref{eq:strong4} at the bottom boundary $\Gamma_\text{b}$. Then, by substituting this velocity field into Eq.~\eqref{eq:strong1}, we obtain an  explicit expression of the forcing term $\bf f$ in Eq.~\eqref{eq:strong1}, a function $p_0$ for the traction-free boundary condition Eq.~\eqref{eq:strong3} at the top boundary $\Gamma_\text{t}$, and a function $\beta$ for the sliding boundary condition Eq.~\eqref{eq:strong5} at the bottom boundary $\Gamma_\text{b}$. 
%
%

Since the shape of the domain is a semicircle in this example, it is not necessary to use normalization layers or balance regularization factors. We use a constant regularization factor $\frac{1}{\varepsilon}=50$ for all boundary conditions. During training, in every step of stochastic gradient descent, we use $5000$ random samples uniformly distributed in the domain $\Omega$ to evaluate Eq.~\eqref{num_lag} and $1000$ uniform random samples on the boundary $\Gamma$ to construct the soft constraints in Eq.~\eqref{eq:strong3}--Eq.~\eqref{eq:strong5}. The learning rate, which is the step length of SGD, is set to be $10^{-3}$. We renew the training data every $200$ steps of learning. The configuration of the neural network is illustrated in Fig.~\ref{DNN_phi}.

\begin{figure}[htbp]
	\centering
	\begin{subfigure}{0.32\textwidth}
		\includegraphics[width = \linewidth]{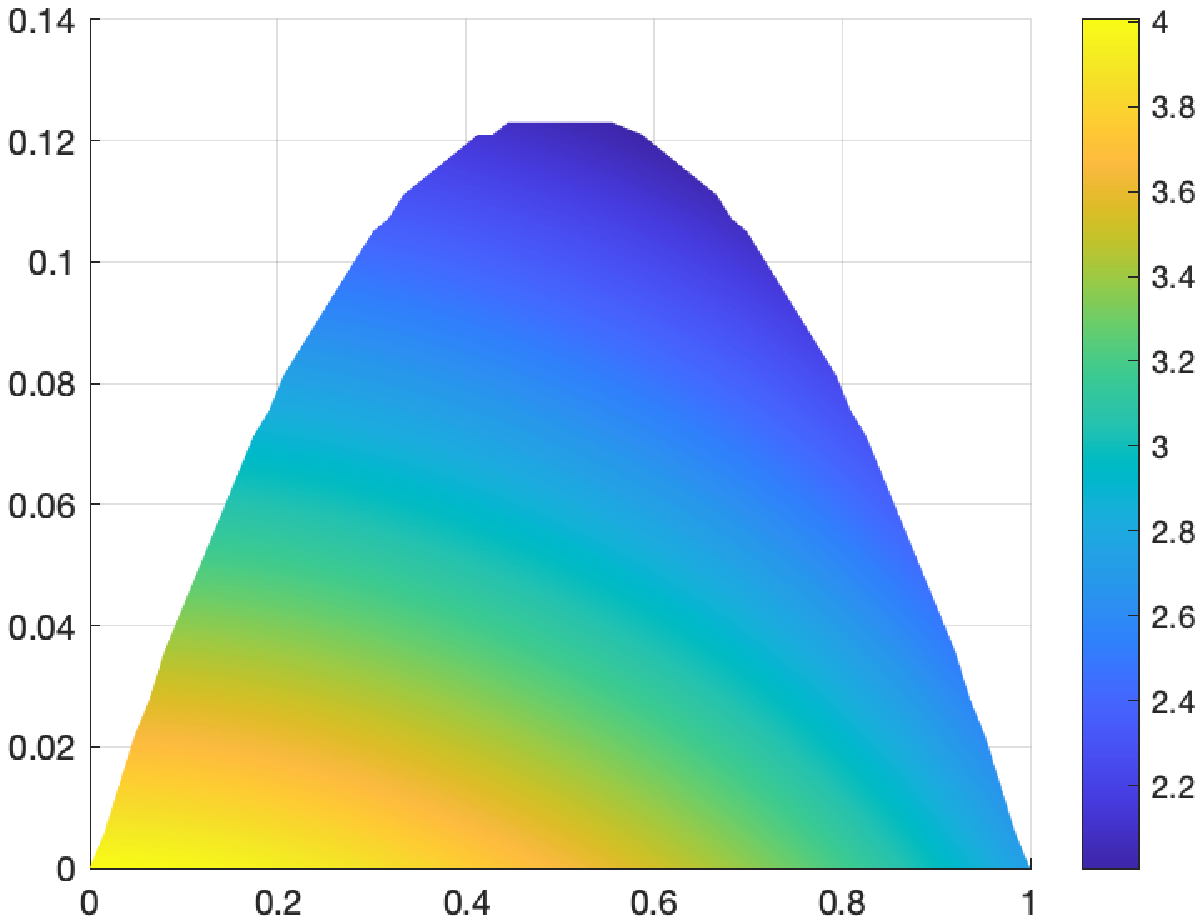}
		\caption{$u$}
	\end{subfigure}
		\begin{subfigure}{0.32\textwidth}
		\includegraphics[width = \linewidth]{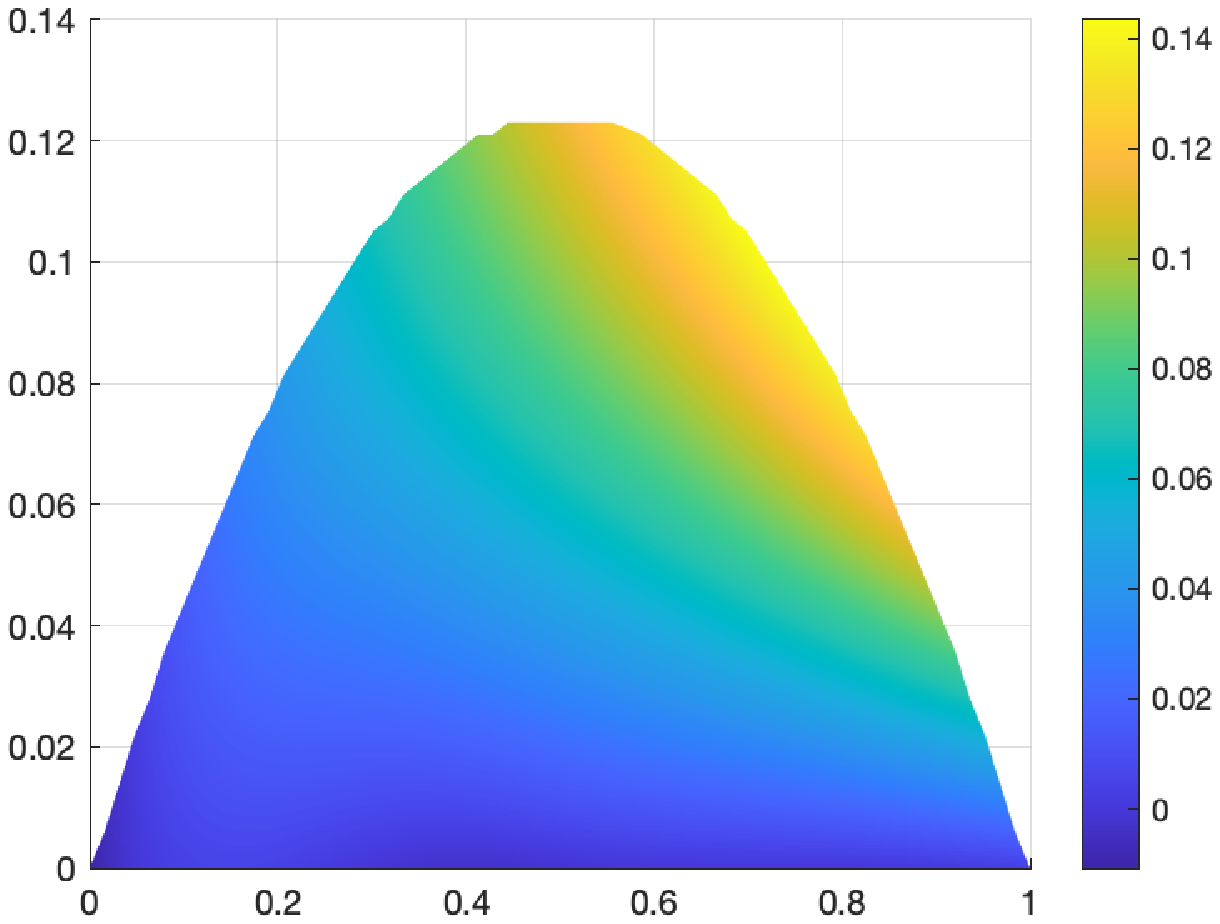}
		\caption{$v$}
	\end{subfigure}
		\begin{subfigure}{0.32\textwidth}
		\includegraphics[width = \linewidth]{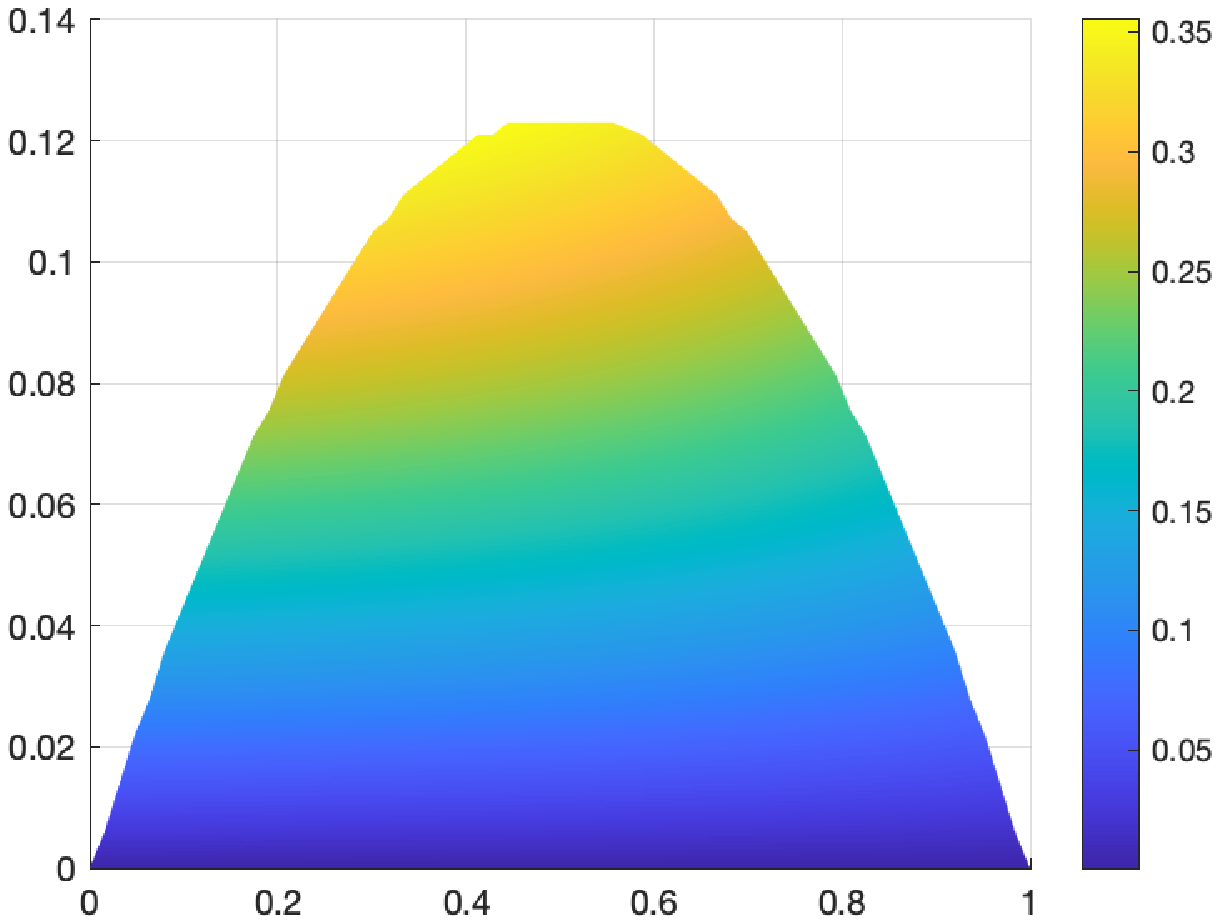}
		\caption{$\phi$ rescaled}
	\end{subfigure}
	\\
		\begin{subfigure}{0.32\textwidth}
		\includegraphics[width = \linewidth]{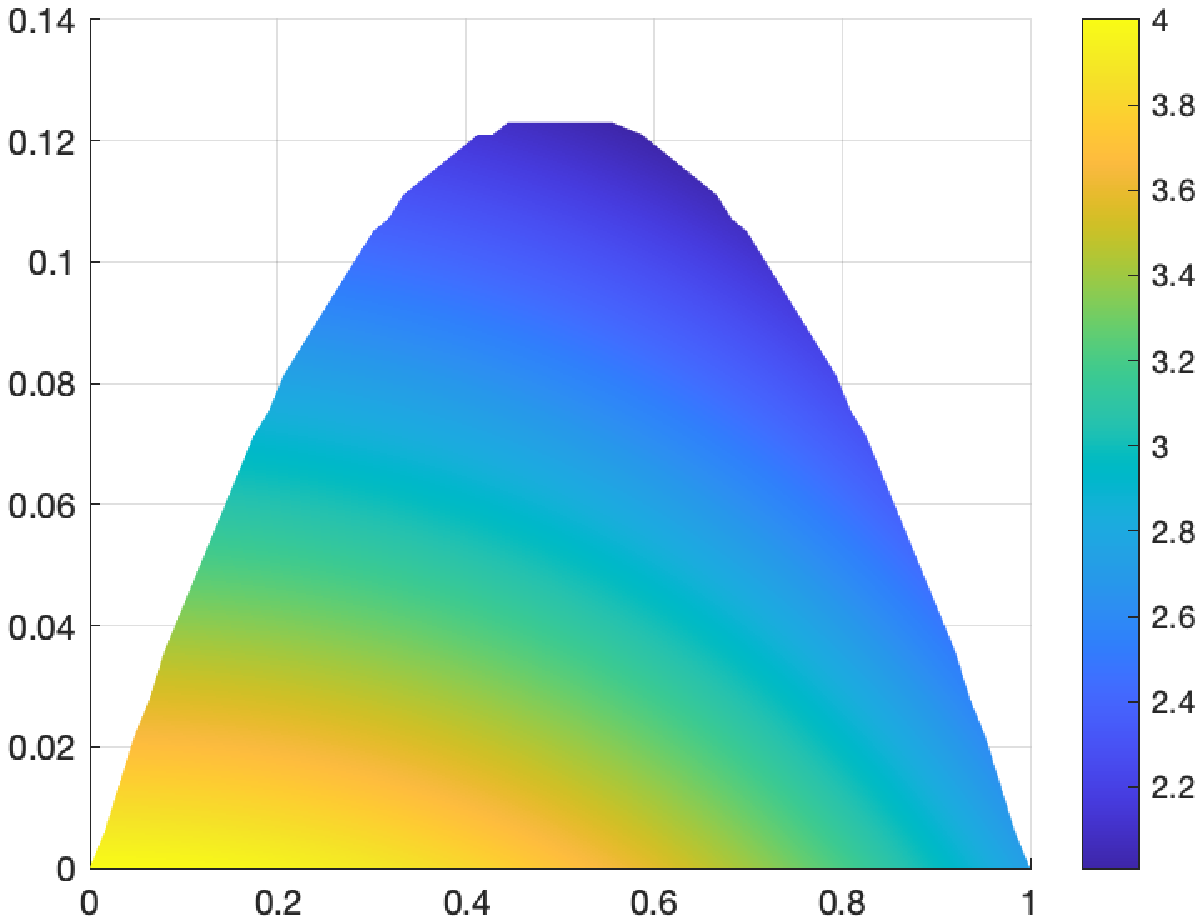}
		\caption{ref: $u$}
	\end{subfigure}
		\begin{subfigure}{0.32\textwidth}
		\includegraphics[width = \linewidth]{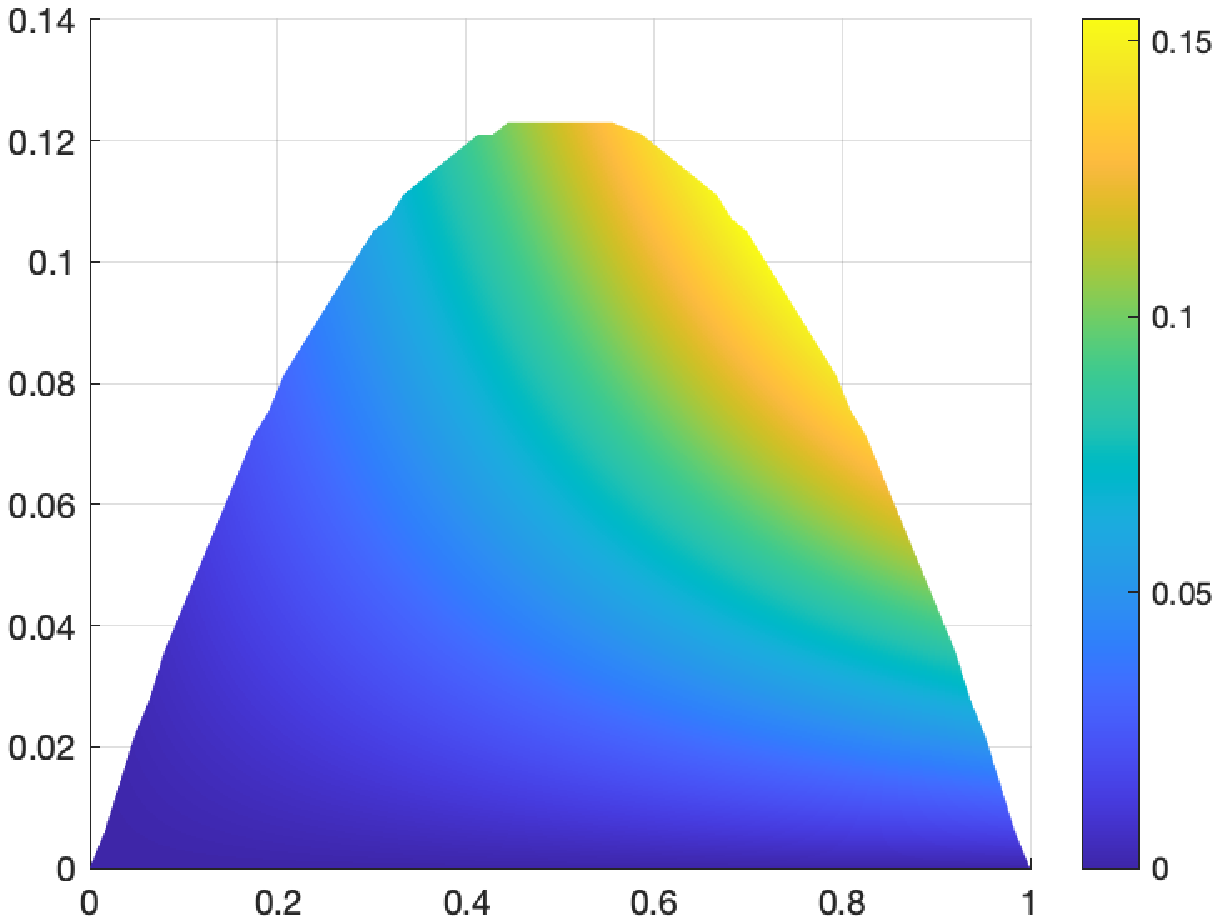}
		\caption{ref $v$}
	\end{subfigure}
		\begin{subfigure}{0.32\textwidth}
		\includegraphics[width = \linewidth]{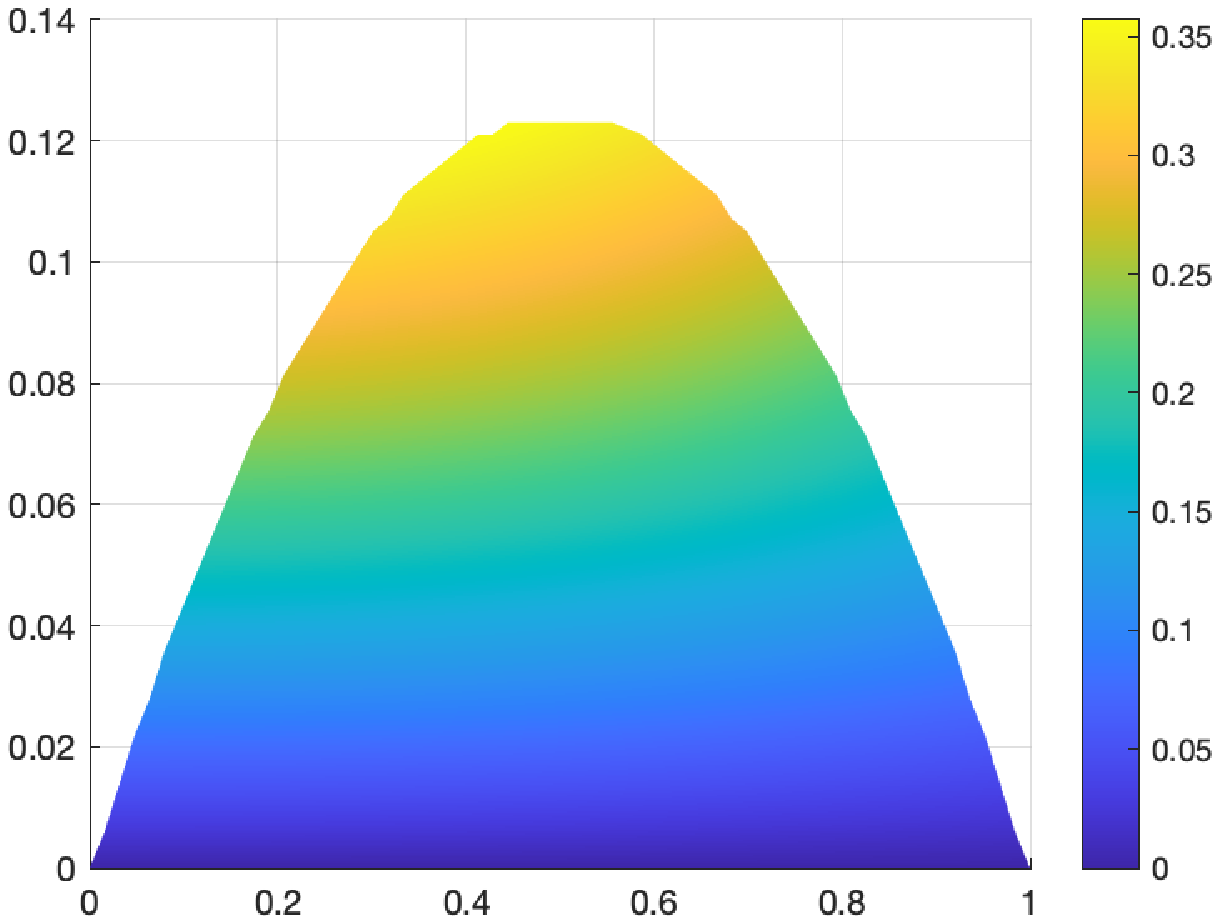}
		\caption{ref $\phi$}
	\end{subfigure}
	\caption{Solution of synthetic model}
	\label{fig:eg1_uvphi}
\end{figure}

In Fig.~\ref{fig:eg1_uvphi}, we compare the results of our proposed method with the ground truth. Since the velocity field is invariant to the potential function $\phi$ up to a constant shift, we shift the output of $\phi$ learned by neural networks to ensure it has the same average as the ground truth in the comparison. For both of the velocity field and the potential function, the neural network offers comparable results with the ground truth.

\begin{figure}
    \centering
    \includegraphics[width=0.45\textwidth]{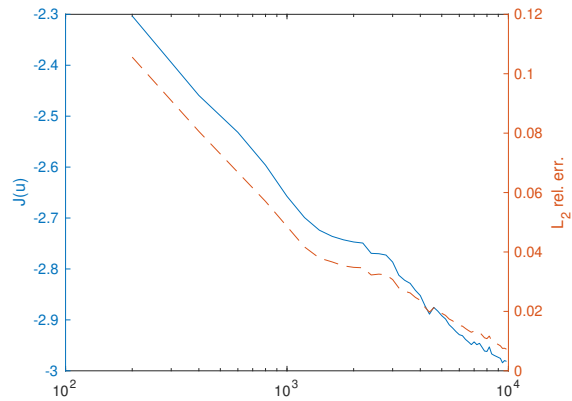}
    \caption{Training loss (Lagrangian) and validation error ($L_2$ relative error) with respect to training steps}
    \label{fig:eg1_logs}
\end{figure}

In Fig.~\ref{fig:eg1_logs} we compare the relative $L_2$ error of the velocity field with the energy functional $\mathcal{E}_\mathrm{s}(\vel)$ in Eq. \eqref{eq:energy_stokes}. We observe that $\mathcal{E}_\mathrm{s}(\vel)$ follows a similar trend to the relative $L_2$ error. The relative $L_2$ error drops to $0.005$ after $10000$ steps of training.

\subsection{The Arolla model}

\begin{figure}
    \centering
    \begin{subfigure}{0.32\textwidth}
		\includegraphics[width = \linewidth]{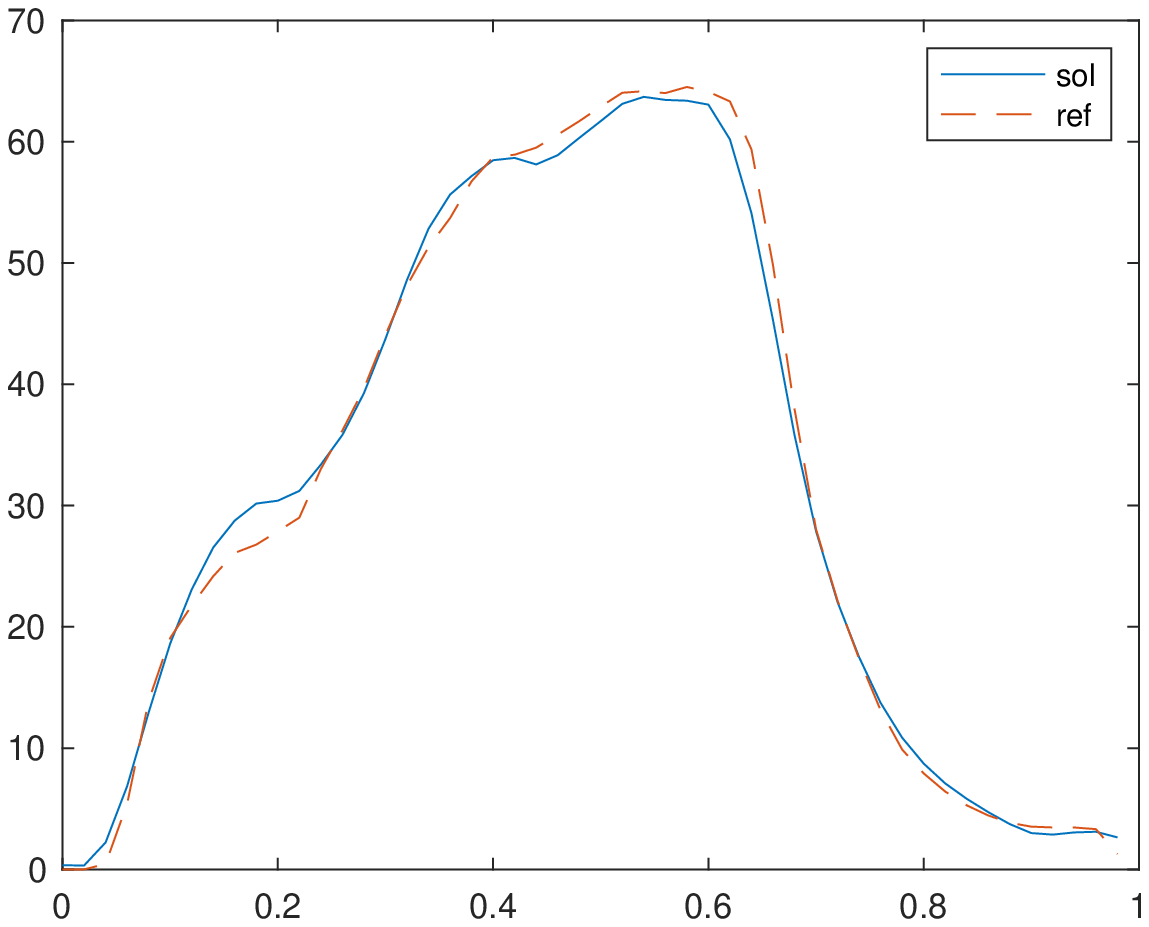}
		\caption{Surface Velocity along surface direction}
	\end{subfigure}
	\begin{subfigure}{0.32\textwidth}
		\includegraphics[width = \linewidth]{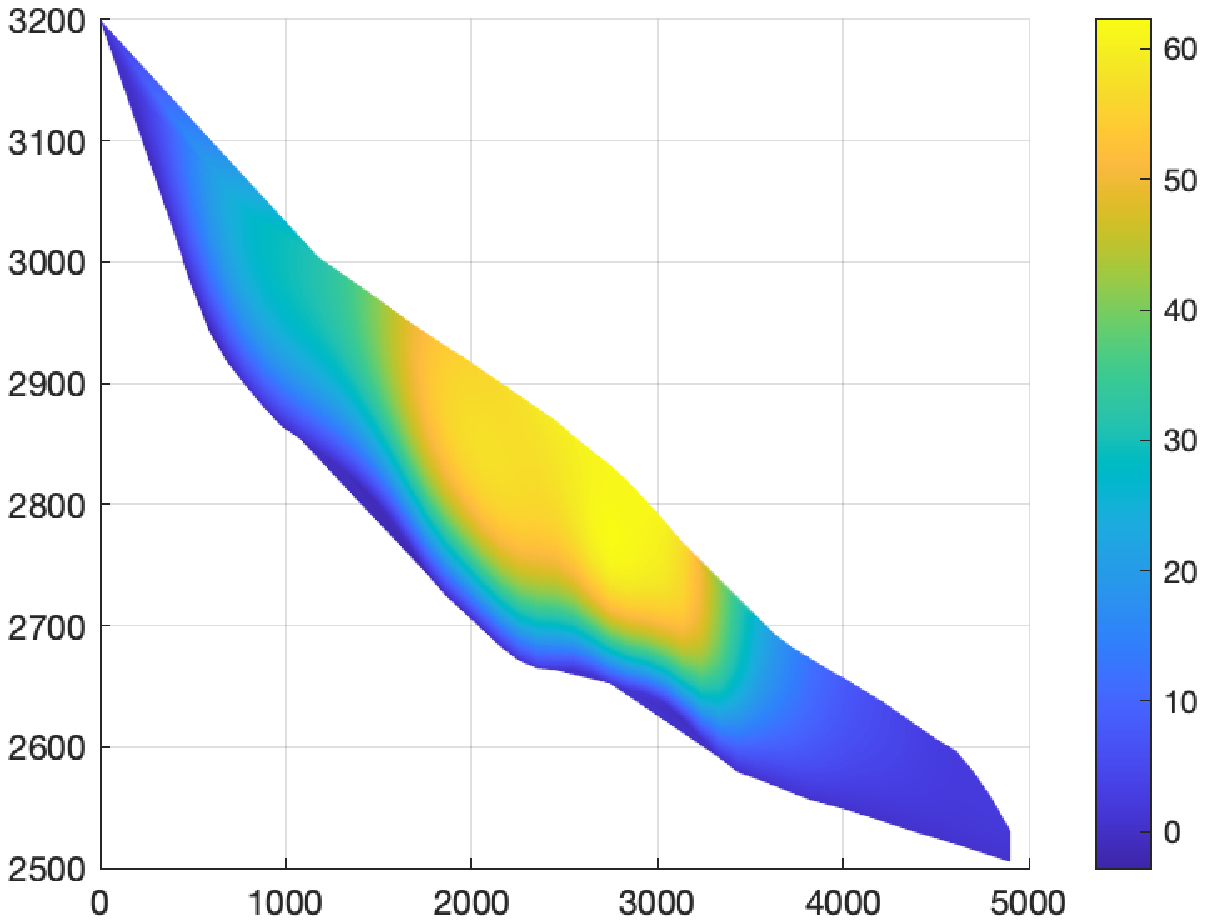}
		\caption{$u$}
	\end{subfigure}
	\begin{subfigure}{0.32\textwidth}
		\includegraphics[width = \linewidth]{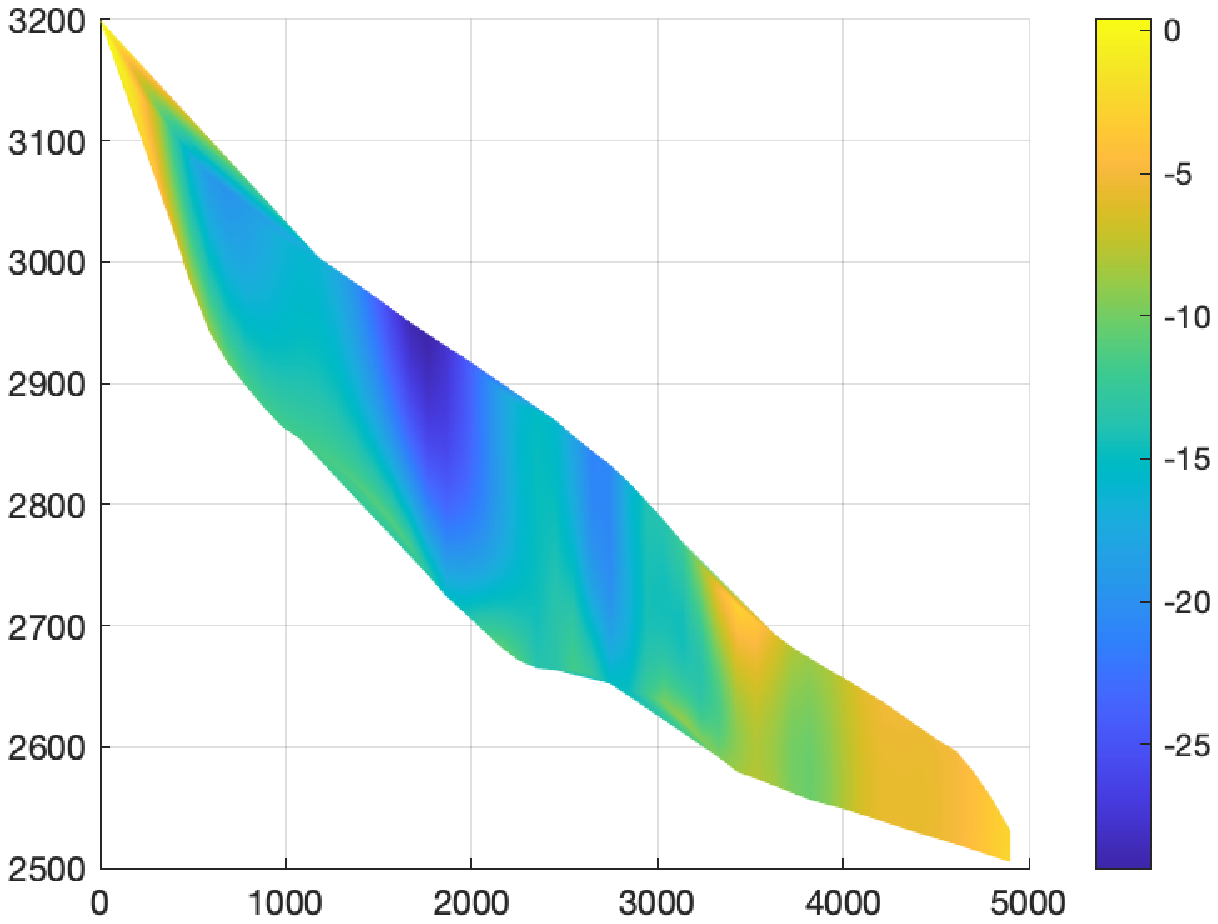}
		\caption{$v$}
	\end{subfigure}
    \caption{Arolla}
    \label{fig:eg2_surface}
\end{figure}

The two-dimensional model of the Arolla glacier is a diagnostic experiment along the central flow-line of a temperate glacier in the European Alps (Haut Glacier d'Arolla, Switzerland), based on earlier experiments by \cite{blatter1998stress,pattyn2002transient}. The domain of this model is enclosed by the longitudinal surface and bedrock profiles of Haut Glacier d'Arolla (as shown in Fig. \ref{fig:eg2_surface} (b) and (c)). This model has been used as a benchmark example in various investigations of the forward modelling and the inverse modelling of glaciers, see \cite{petra2014computational} for some notable examples. 

We consider two experiments in this example. The first experiment follows the setup of \cite[Section 3.5]{pattyn2008benchmark}, in which a no-sliding bottom boundary with $\beta = \infty$ is considered. In the second experiment, we use a sliding bottom boundary with the function 
\[
\exp(\beta) = \left\{ \begin{array}{ll} 
1000 + 1000 \sin(\frac{2 \pi x}{5000}) & 0 \leq x < 3750\\
1000 (16 - \frac{x}{250}) & 3750 \leq x < 4000\\
1000 & 4000 \leq x < 5000
\end{array}\right.
\]
which is the same as the work of \cite[Section 4.2]{petra2014computational}.

\begin{figure}
    \centering
 \begin{subfigure}{0.32\textwidth}
		\includegraphics[width = \linewidth]{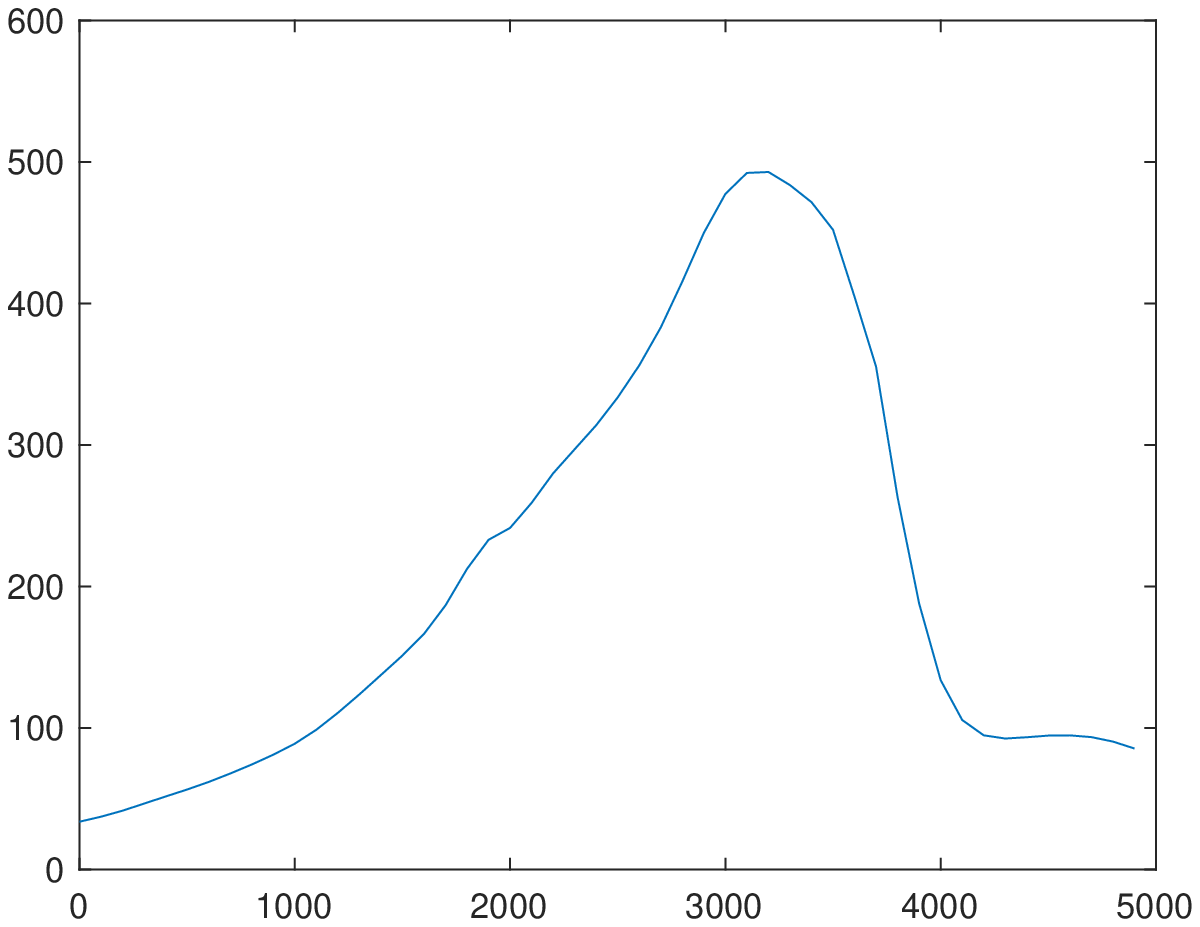}
		\caption{Surface Velocity along surface direction}
	\end{subfigure}
	\begin{subfigure}{0.32\textwidth}
		\includegraphics[width = \linewidth]{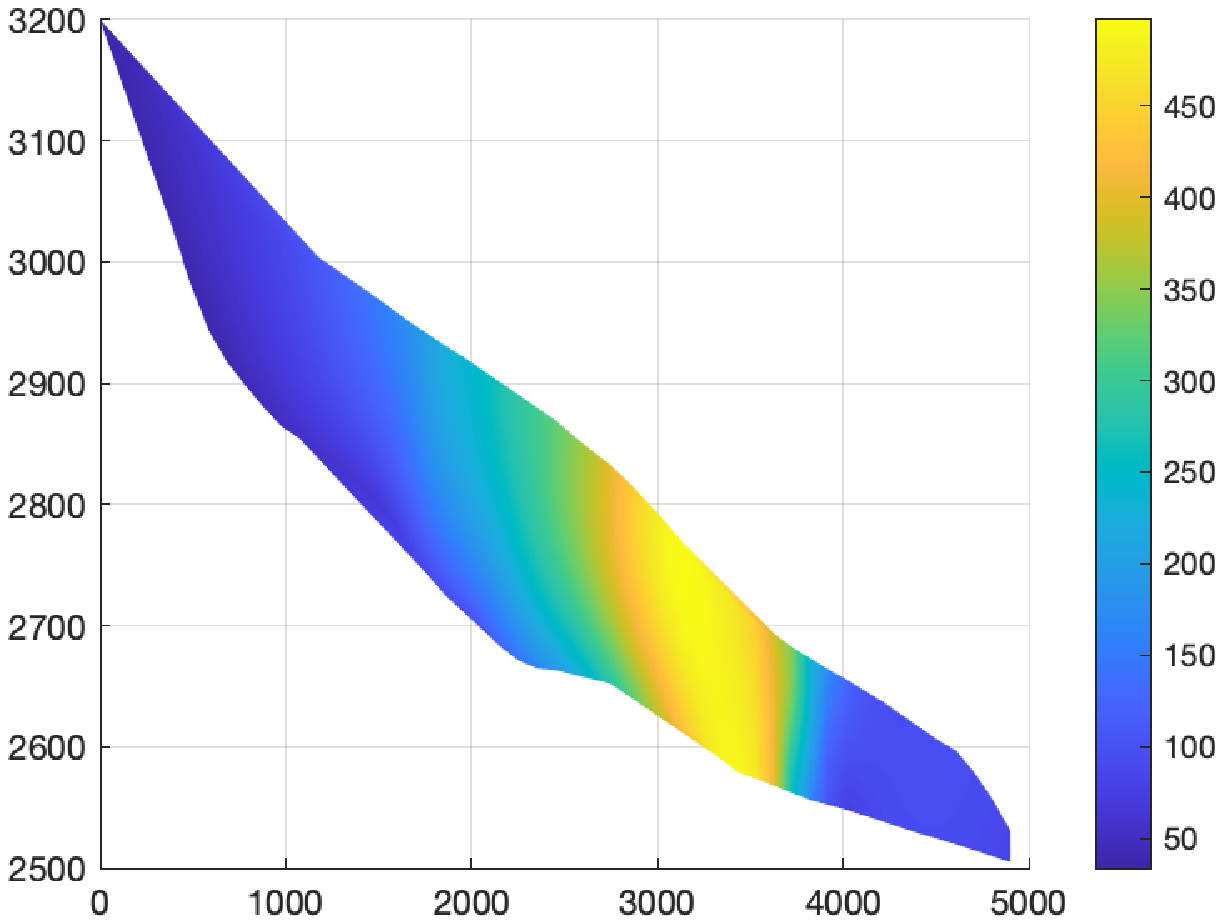}
		\caption{$u$}
	\end{subfigure}
	\begin{subfigure}{0.32\textwidth}
		\includegraphics[width = \linewidth]{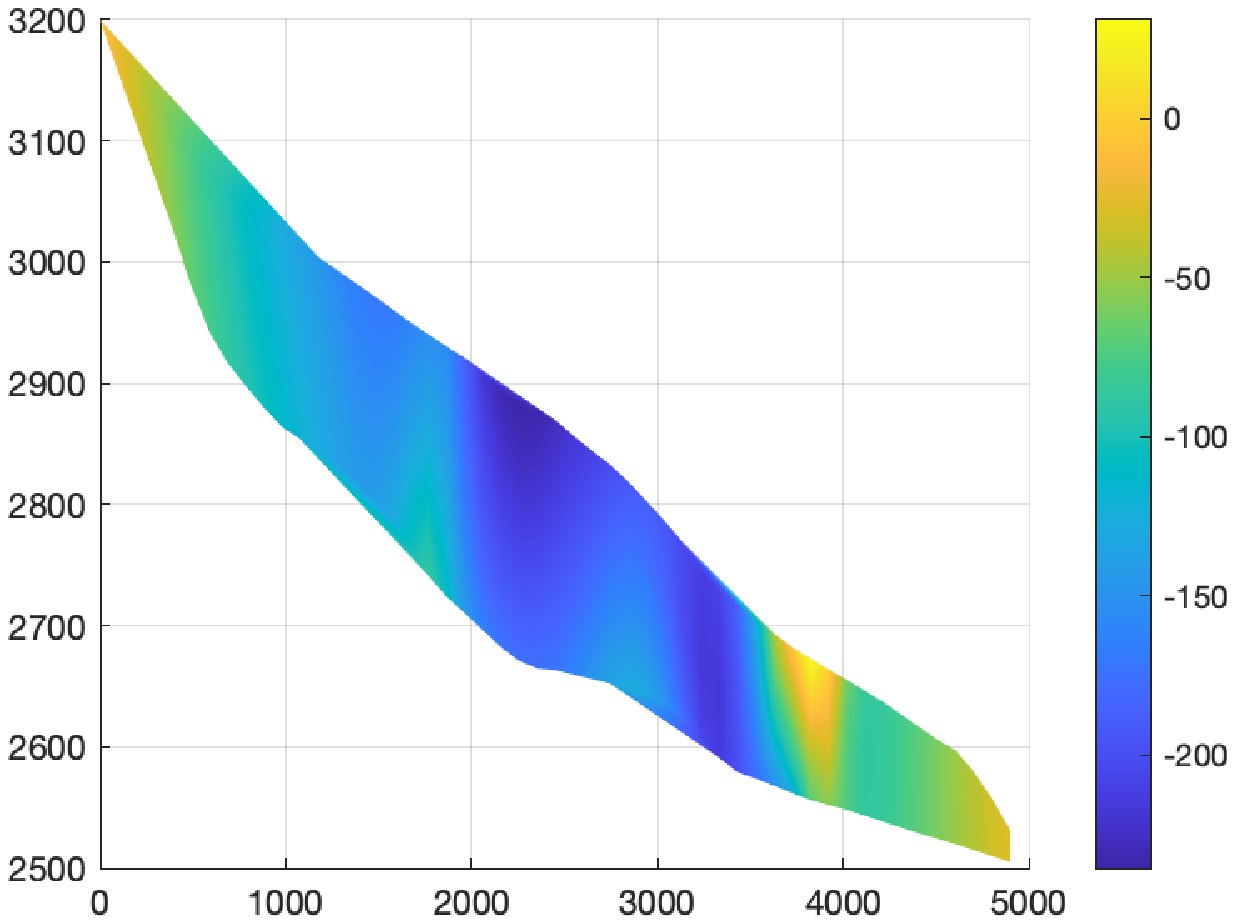}
		\caption{$v$}
	\end{subfigure}    \caption{Noemi's paper}
    \label{fig:eg3_surface}
\end{figure}

\subsection{A 3D box model}

\begin{figure}
    \centering
 \begin{subfigure}[b]{0.45\textwidth}
		\includegraphics[width = \linewidth]{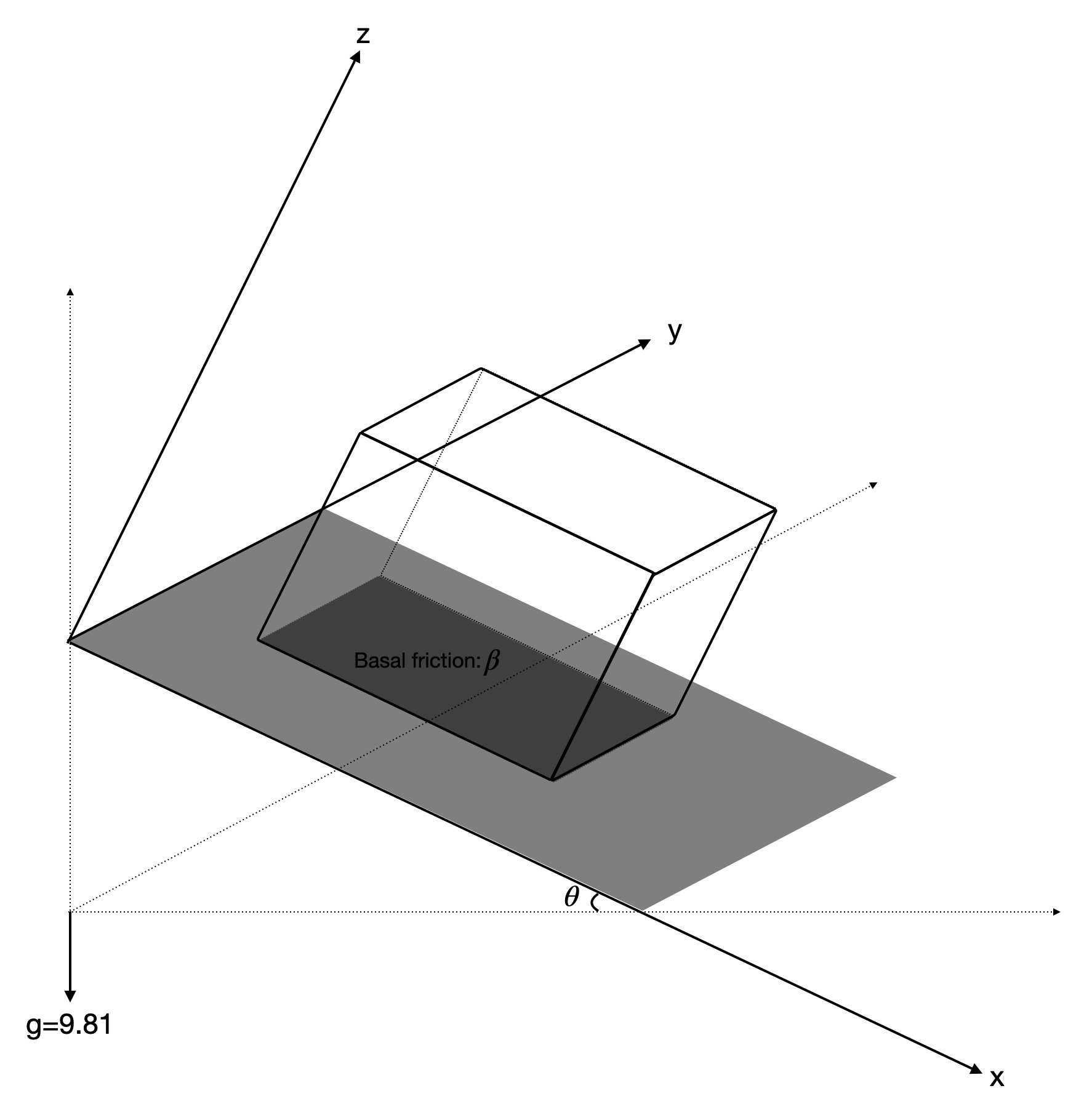}
		\caption{Illustration of coordinate change}
	\end{subfigure}
	\begin{subfigure}[b]{0.53\textwidth}
		\includegraphics[width = \linewidth]{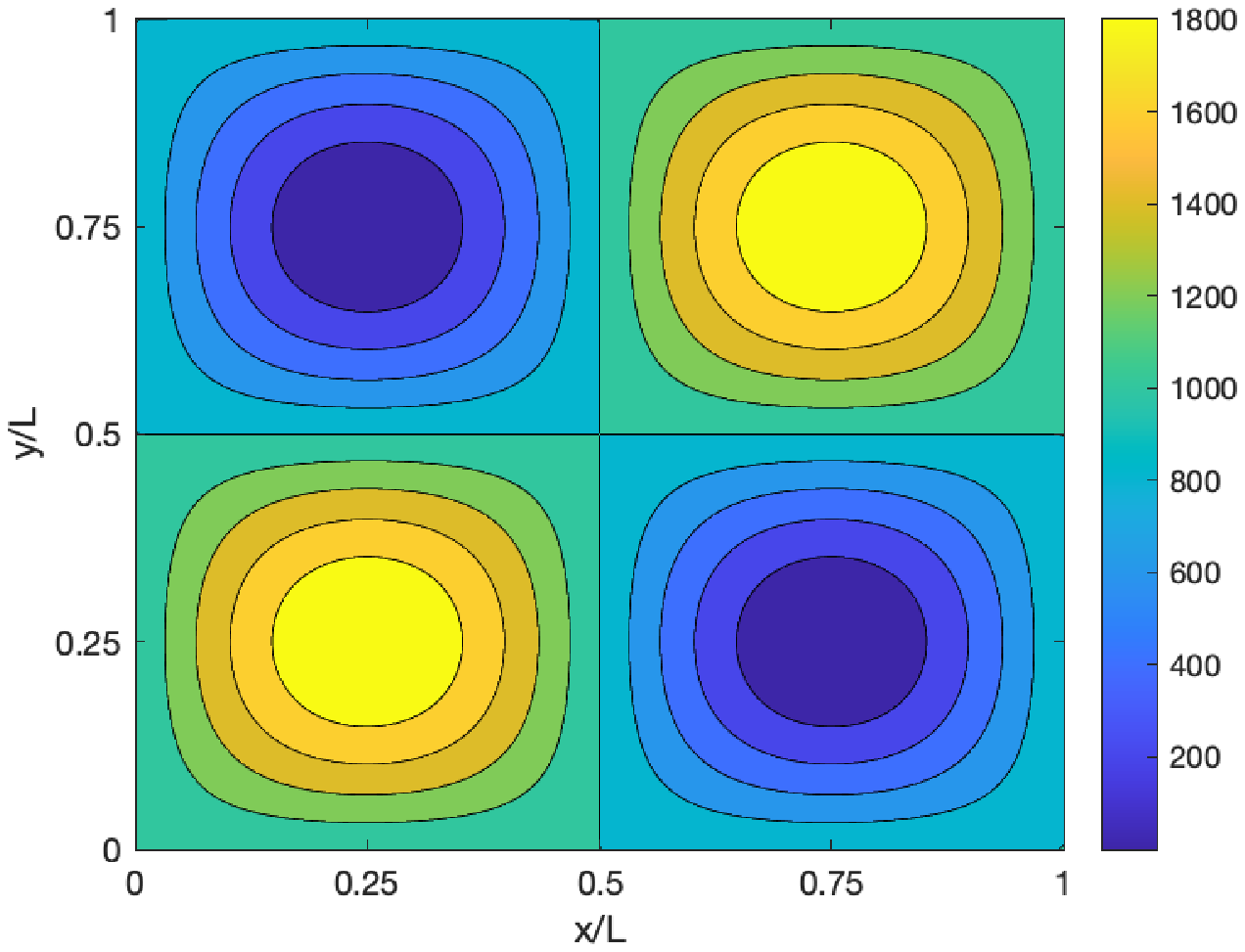}
		\caption{Value of Basal Coefficients $\beta$}
	\end{subfigure}
   \caption{Setting of box Model}
    \label{fig:eg4_illustration}
\end{figure}

\begin{figure}
    \centering
    \includegraphics[width=0.7\textwidth]{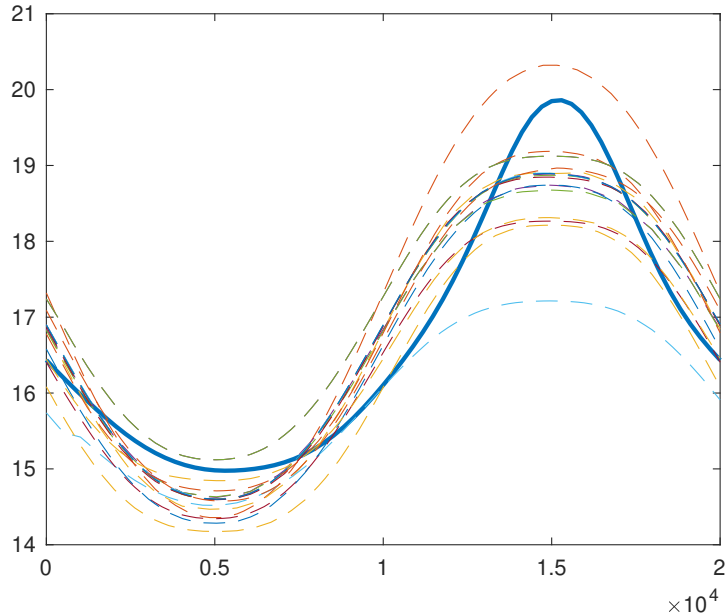}
    \caption{Surface velocity, Benchmark paper, experiment C, 20km. Solid line: Our prediction; Dashed lines: various prediction in Benchmark paper}
    \label{fig:eg4_surface}
\end{figure}

We also consider the three dimensional box model presented in \cite[Section 3]{pattyn2008benchmark}. In this model, a slab of ice sliding down a sloping bed with a constant incline angle $\alpha = 0.1$ degree is considered. The ice slab is enclosed in a box-shaped domain $[0,L]\times[0,L]\times [0,H]$, where $L = 20 \times 10^3$ metres, and $H = 10^3$ meters. After a change of coordinate that align the $x$-axis with the sliding surface (see \ref{fig:eg4_illustration}(a)), the vector $\rho \grav$ is given as
\[
\rho \grav = 910 \times 9.81 \times [ \sin\theta, 0, - \cos\theta ] \, \text{Pa}, 
\]
where $\theta = \pi/1800$. 
In this example, periodic boundary conditions are applied to boundaries in the horizontal directions, the traction-free boundary conditions in Eq.~\eqref{eq:strong3} is applied to the top boundary, and the sliding boundary condition in Eq.~\eqref{eq:strong4} and Eq.~\eqref{eq:strong5} is prescribed at the bottom boundary. The basal sliding coefficient is defined as
\[
\beta(x, y) = 1000 + 1000 \sin\Big(\frac{2\pi x}{L}\Big) \sin\Big(\frac{2\pi y}{L}\Big).
\]
In Fig.\ref{fig:eg4_illustration}(b) we plot the value of basal friction coefficient $\beta$. Empirically speaking, the surface speed becomes larger when $\beta$ is small. The ground truth solution is unknown. In Fig.\ref{fig:eg4_surface} we plot the surface velocity against $y$ direction at the slice $x=\frac{L}{4}$. In addition we compare our prediction with various result in \cite{pattyn2008benchmark} (experiment C, $L=20$km).

\section{Conclusions}
\noindent
In this paper, we investigate deep-learning methods to simulate non-Newtonian ice flow models. By formulating the model problem into a variational form and constructing a divergence-free solution space using DNN, we convert the solution of the nonlinear PDE into the solution of an optimization problem. 
Despite the rather high-dimensionality of the parameter space of the DNN, the solution can still be efficiently obtained by the SGD method. %
In this framework, once we are able to sample from the problem domain and its boundary, no special treatment is needed to handle irregular boundaries of the problem domain. Therefore, the proposed method is easy to implement and mesh-free. To address the application in real-world computation, we introduce normalizing layers and adaptive boundary penalties in the configuration of our network.  Finally, we present numerical experiments to demonstrate the performance of the proposed method. Specifically, we use the DNN method to solve ice flow model for the 2D Arrola Glacier and a 3D box model with real-world scaling.  Our numerical experiments demonstrate that the DNN method provides satisfactory simulation results for glacier flows. We expect this method can be applied to a more general class of problems in non-Newtonian mechanics that satisfies certain variational principles. 


\section*{Acknowledgements}
\noindent
The research of T. Cui is supported by the Australian Research Council under the grant DP21010309. The research of Z. Zhang is supported by Hong Kong RGC grant (Projects 17300318 and 17307921), National Natural Science Foundation of China  (Project 12171406), Seed Funding Programme for Basic Research (HKU) and Seed Funding for Strategic Interdisciplinary Research Scheme 2021/22 (HKU).




\bibliographystyle{siam}
\bibliography{ZWpaper_DeepLearningPDEDiscCoef}

\begin{thebibliography}{10}

\bibitem{BaoZhou_20b}
{\sc G.~Bao, X.~Ye, Y.~Zang, and H.~Zhou}, {\em Numerical solution of inverse
  problems by weak adversarial networks}, Inverse Problems, 36(11) (2020),
  p.~115003.

\bibitem{Brenner_18}
{\sc Y.~Bar-Sinai, S.~Hoyer, J.~Hickey, and M.~Brenner}, {\em Learning
  data-driven discretizations of {PDE}s}, Bulletin of the American Physical
  Society, 63 (2018).

\bibitem{blatter1998stress}
{\sc H.~Blatter, G.~K. Clarke, and J.~Colinge}, {\em Stress and velocity fields
  in glaciers: Part ii. sliding and basal stress distribution}, Journal of
  Glaciology, 44 (1998), pp.~457--466.

\bibitem{bottou2010large}
{\sc L.~Bottou}, {\em Large-scale machine learning with stochastic gradient
  descent}, in Proceedings of COMPSTAT'2010, Springer, 2010, pp.~177--186.

\bibitem{brown2010efficient}
{\sc J.~Brown}, {\em Efficient nonlinear solvers for nodal high-order finite
  elements in 3d}, Journal of Scientific Computing, 45 (2010), pp.~48--63.

\bibitem{brown2013achieving}
{\sc J.~Brown, B.~Smith, and A.~Ahmadia}, {\em Achieving textbook multigrid
  efficiency for hydrostatic ice sheet flow}, SIAM Journal on Scientific
  Computing, 35 (2013), pp.~B359--B375.

\bibitem{Cai_21}
{\sc Z.~Cai, J.~Chen, and M.~Liu}, {\em Least-squares {ReLU} neural network
  ({LSNN}) method for linear advection-reaction equation}, Journal of
  Computational Physics,  (2021), p.~110514.

\bibitem{Cai_20}
{\sc Z.~Cai, J.~Chen, M.~Liu, and X.~Liu}, {\em Deep least-squares methods: An
  unsupervised learning-based numerical method for solving elliptic {PDE}s},
  Journal of Computational Physics, 420 (2020), p.~109707.

\bibitem{cohen2016expressive}
{\sc N.~Cohen, O.~Sharir, and A.~Shashua}, {\em On the expressive power of deep
  learning: {A} tensor analysis}, in Conference on Learning Theory, 2016,
  pp.~698--728.

\bibitem{cybenko1989approximation}
{\sc G.~Cybenko}, {\em Approximation by superpositions of a sigmoidal
  function}, Mathematics of control, signals and systems, 2 (1989),
  pp.~303--314.

\bibitem{dukowicz2010consistent}
{\sc J.~K. Dukowicz, S.~F. Price, and W.~H. Lipscomb}, {\em Consistent
  approximations and boundary conditions for ice-sheet dynamics from a
  principle of least action}, Journal of Glaciology, 56 (2010), pp.~480--496.

\bibitem{weinan2018deep}
{\sc W.~E and B.~Yu}, {\em The deep {R}itz method: A deep learning-based
  numerical algorithm for solving variational problems}, Communications in
  Mathematics and Statistics, 6 (2018), pp.~1--12.

\bibitem{ellacott1994aspects}
{\sc S.~Ellacott}, {\em Aspects of the numerical analysis of neural networks},
  Acta Numerica, 3 (1994), pp.~145--202.

\bibitem{elman2014finite}
{\sc H.~C. Elman, D.~J. Silvester, and A.~J. Wathen}, {\em Finite elements and
  fast iterative solvers: with applications in incompressible fluid dynamics},
  Numerical Mathematics and Scie, 2014.

\bibitem{fraters2019efficient}
{\sc M.~R. Fraters, W.~Bangerth, C.~Thieulot, A.~Glerum, and W.~Spakman}, {\em
  Efficient and practical newton solvers for non-linear stokes systems in
  geodynamic problems}, Geophysical Journal International, 218 (2019),
  pp.~873--894.

\bibitem{goodfellow2016deep}
{\sc I.~Goodfellow, Y.~Bengio, A.~Courville, and Y.~Bengio}, {\em Deep
  learning}, vol.~1, MIT press Cambridge, 2016.

\bibitem{han2018solving}
{\sc J.~Han, A.~Jentzen, and W.~E}, {\em Solving high-dimensional partial
  differential equations using deep learning}, Proceedings of the National
  Academy of Sciences, 115 (2018), pp.~8505--8510.

\bibitem{JinchaoXu:2018}
{\sc J.~He, L.~Li, J.~Xu, and C.~Zheng}, {\em Relu deep neural networks and
  linear finite elements}, Journal of Computational Mathematics, 38 (2020),
  pp.~502--527.

\bibitem{Xu_mgnet}
{\sc J.~He and J.~Xu}, {\em Mgnet: A unified framework of multigrid and
  convolutional neural network}, Science {C}hina {M}athematics, 62 (2019),
  pp.~1331--1354.

\bibitem{heuveline2007inf}
{\sc V.~Heuveline and F.~Schieweck}, {\em On the inf-sup condition for higher
  order mixed fem on meshes with hanging nodes}, ESAIM: Mathematical Modelling
  and Numerical Analysis, 41 (2007), pp.~1--20.

\bibitem{hornik1989multilayer}
{\sc K.~Hornik, M.~Stinchcombe, and H.~White}, {\em Multilayer feedforward
  networks are universal approximators}, Neural networks, 2 (1989),
  pp.~359--366.

\bibitem{hutter2017theoretical}
{\sc K.~Hutter}, {\em Theoretical glaciology: material science of ice and the
  mechanics of glaciers and ice sheets}, vol.~1, Springer, 2017.

\bibitem{isaac2015solution}
{\sc T.~Isaac, G.~Stadler, and O.~Ghattas}, {\em Solution of nonlinear stokes
  equations discretized by high-order finite elements on nonconforming and
  anisotropic meshes, with application to ice sheet dynamics}, SIAM Journal on
  Scientific Computing, 37 (2015), pp.~B804--B833.

\bibitem{jingrun_deepritz}
{\sc Jingrun, Chen, , 8405, , J.~Chen, Rui, Du, , 8406, , R.~Du, Keke, Wu, ,
  8407, , and K.~Wu}, {\em A comparison study of deep galerkin method and deep
  ritz method for elliptic problems with different boundary conditions},
  Communications in Mathematical Research, 36 (2020), pp.~354--376.

\bibitem{karumuri2019simulator}
{\sc S.~Karumuri, R.~Tripathy, I.~Bilionis, and J.~Panchal}, {\em
  Simulator-free solution of high-dimensional stochastic elliptic partial
  differential equations using deep neural networks}, Journal of Computational
  Physics, 404 (2020), p.~109120.

\bibitem{khoo2017solving}
{\sc Y.~Khoo, J.~Lu, and L.~Ying}, {\em Solving parametric {PDE} problems with
  artificial neural networks}, European Journal of Applied Mathematics, Special
  Issue 3: Connections between Deep Learning and Partial Differential
  Equations, 32 (2021), pp.~421--435.

\bibitem{kingma2014adam}
{\sc D.~Kingma and J.~Ba}, {\em Adam: {A} method for stochastic optimization},
  arXiv:1412.6980,  (2014).

\bibitem{Lagaris1998}
{\sc I.~Lagaris, A.~Likas, and D.~Fotiadis}, {\em Artificial neural networks
  for solving ordinary and partial differential equations}, IEEE Trans. Neural
  Netw., 9 (1998), pp.~987--1000.

\bibitem{lecun2015deep}
{\sc Y.~LeCun, Y.~Bengio, and G.~Hinton}, {\em Deep learning}, Nature, 521
  (2015), p.~436.

\bibitem{LeeH1990}
{\sc H.~Lee and I.~S. Kang}, {\em Neural algorithm for solving differential
  equations}, Journal of Computational Physics, 91 (1990), pp.~110--131.

\bibitem{li2020fourier}
{\sc Z.~Li, N.~Kovachki, K.~Azizzadenesheli, B.~Liu, K.~Bhattacharya,
  A.~Stuart, and A.~Anandkumar}, {\em Fourier neural operator for parametric
  partial differential equations}, arXiv preprint arXiv:2010.08895,  (2020).

\bibitem{Dongb_19}
{\sc Z.~Long, Y.~Lu, and B.~Dong}, {\em {PDE-Net} 2.0: {L}earning {PDE}s from
  data with a numeric-symbolic hybrid deep network}, Journal of Computational
  Physics, 399 (2019), p.~108925.

\bibitem{Dongb_18}
{\sc Z.~Long, Y.~Lu, X.~Ma, and B.~Dong}, {\em {PDE}-{N}et: {L}earning {PDE}s
  from data}, International Conference on Machine Learning,  (2018),
  pp.~3208--3216.

\bibitem{Shen_21a}
{\sc J.~Lu, Z.~Shen, H.~Yang, and S.~Zhang}, {\em Deep network approximation
  for smooth functions}, SIAM Journal on Mathematical Analysis, 53 (2021),
  pp.~5465--5506.

\bibitem{lu2019deeponet}
{\sc L.~Lu, P.~Jin, and G.~Karniadakis}, {\em Deeponet: {L}earning nonlinear
  operators for identifying differential equations based on the universal
  approximation theorem of operators}, arXiv:1910.03193,  (2019).

\bibitem{may2015scalable}
{\sc D.~A. May, J.~Brown, and L.~Le~Pourhiet}, {\em A scalable, matrix-free
  multigrid preconditioner for finite element discretizations of heterogeneous
  stokes flow}, Computer methods in applied mechanics and engineering, 290
  (2015), pp.~496--523.

\bibitem{mckenzie1984generation}
{\sc D.~McKenzie}, {\em The generation and compaction of partially molten
  rock}, Journal of petrology, 25 (1984), pp.~713--765.

\bibitem{Fernandez1994}
{\sc A.~Meade and A.~Fernandez}, {\em The numerical solution of linear ordinary
  differential equations by feedforward neural networks}, Math. Comput. Model.,
  19 (1994), pp.~1--25.

\bibitem{QiangDU:2018}
{\sc H.~Montanelli and Q.~Du}, {\em New error bounds for deep {R}e{L}{U}
  networks using sparse grids}, SIAM Journal on Mathematics of Data Science, 1
  (2019), pp.~78--92.

\bibitem{paterson1994physics}
{\sc W.~S.~B. Paterson}, {\em Physics of glaciers}, Butterworth-Heinemann,
  1994.

\bibitem{pattyn2002transient}
{\sc F.~Pattyn}, {\em Transient glacier response with a higher-order numerical
  ice-flow model}, Journal of Glaciology, 48 (2002), pp.~467--477.

\bibitem{pattyn2008benchmark}
{\sc F.~Pattyn, L.~Perichon, A.~Aschwanden, B.~Breuer, B.~De~Smedt,
  O.~Gagliardini, G.~H. Gudmundsson, R.~C. Hindmarsh, A.~Hubbard, J.~V.
  Johnson, et~al.}, {\em Benchmark experiments for higher-order and full-stokes
  ice sheet models (ismip--hom)}, The Cryosphere, 2 (2008), pp.~95--108.

\bibitem{petra2014computational}
{\sc N.~Petra, J.~Martin, G.~Stadler, and O.~Ghattas}, {\em A computational
  framework for infinite-dimensional bayesian inverse problems, part ii:
  Stochastic newton mcmc with application to ice sheet flow inverse problems},
  SIAM Journal on Scientific Computing, 36 (2014), pp.~A1525--A1555.

\bibitem{petra2012inexact}
{\sc N.~Petra, H.~Zhu, G.~Stadler, T.~J. Hughes, and O.~Ghattas}, {\em An
  inexact gauss-newton method for inversion of basal sliding and rheology
  parameters in a nonlinear stokes ice sheet model}, Journal of Glaciology, 58
  (2012), pp.~889--903.

\bibitem{pinkus1999approximation}
{\sc A.~Pinkus}, {\em Approximation theory of the {M}{L}{P} model in neural
  networks}, Acta numerica, 8 (1999), pp.~143--195.

\bibitem{Xiu_19}
{\sc T.~Qin, K.~Wu, and D.~Xiu}, {\em Data driven governing equations
  approximation using deep neural networks}, Journal of Computational Physics,
  395 (2019), pp.~620--635.

\bibitem{raissi2019physics}
{\sc M.~Raissi, P.~Perdikaris, and G.~Karniadakis}, {\em Physics-informed
  neural networks: {A} deep learning framework for solving forward and inverse
  problems involving nonlinear partial differential equations}, Journal of
  Computational Physics, 378 (2019), pp.~686--707.

\bibitem{rudi2015extreme}
{\sc J.~Rudi, A.~C.~I. Malossi, T.~Isaac, G.~Stadler, M.~Gurnis, P.~W. Staar,
  Y.~Ineichen, C.~Bekas, A.~Curioni, and O.~Ghattas}, {\em An extreme-scale
  implicit solver for complex pdes: highly heterogeneous flow in earth's
  mantle}, in Proceedings of the international conference for high performance
  computing, networking, storage and analysis, 2015, pp.~1--12.

\bibitem{Kutz_19}
{\sc S.~Rudy, J.~Kutz, and S.~Brunton}, {\em Deep learning of dynamics and
  signal-noise decomposition with time-stepping constraints}, Journal of
  Computational Physics, 396 (2019), pp.~483--506.

\bibitem{schubert2001mantle}
{\sc G.~Schubert, D.~L. Turcotte, and P.~Olson}, {\em Mantle convection in the
  Earth and planets}, Cambridge University Press, 2001.

\bibitem{schwab1999mixed}
{\sc C.~Schwab and M.~Suri}, {\em Mixed hp finite element methods for stokes
  and non-newtonian flow}, Computer methods in applied mechanics and
  engineering, 175 (1999), pp.~217--241.

\bibitem{schwab2017deep}
{\sc C.~Schwab and J.~Zech}, {\em Deep {L}earning in {H}igh {D}imension},
  Research Report, vol. 2017 (2017).

\bibitem{Shen_21b}
{\sc Z.~Shen, H.~Yang, and S.~Zhang}, {\em Deep network with approximation
  error being reciprocal of width to power of square root of depth}, Neural
  Computation, 33 (2021), pp.~1005--1036.

\bibitem{sirignano2018dgm}
{\sc J.~Sirignano and K.~Spiliopoulos}, {\em {DGM}: {A} deep learning algorithm
  for solving partial differential equations}, Journal of Computational
  Physics, 375 (2018), pp.~1339--1364.

\bibitem{stenberg1996mixed}
{\sc R.~Stenberg and M.~Suri}, {\em Mixed $ hp $ finite element methods for
  problems in elasticity and stokes flow}, Numerische Mathematik, 72 (1996),
  pp.~367--389.

\bibitem{Yalchin2018deep}
{\sc Y.~Wang, S.~Cheung, E.~Chung, Y.~Efendiev, and M.~Wang}, {\em Deep
  multiscale model learning}, Journal of Computational Physics, 406 (2020),
  p.~109071.

\bibitem{wang2020mesh}
{\sc Z.~Wang and Z.~Zhang}, {\em A mesh-free method for interface problems
  using the deep learning approach}, Journal of Computational Physics, 400
  (2020), p.~108963.

\bibitem{Xiu_20}
{\sc K.~Wu and D.~Xiu}, {\em Data-driven deep learning of partial differential
  equations in modal space}, Journal of Computational Physics,  (2020),
  p.~109307.

\bibitem{Karn2021}
{\sc L.~Yang, X.~Meng, and G.~Karniadakis}, {\em B-{PINNS}: Bayesian
  physics-informed neural networks for forward and inverse {PDE} problems with
  noisy data}, J. Comp. Physics, 425 (2021), p.~109913.

\bibitem{yarotsky2017error}
{\sc D.~Yarotsky}, {\em Error bounds for approximations with deep {R}e{L}{U}
  networks}, Neural Networks, 94 (2017), pp.~103--114.

\bibitem{BaoZhou_20a}
{\sc Y.~Zang, G.~Bao, X.~Ye, and H.~Zhou}, {\em Weak adversarial networks for
  high-dimensional partial differential equations}, Journal of Computational
  Physics, 411 (2020), p.~109409.

\bibitem{ZabarasZhu:2018}
{\sc Y.~Zhu and N.~Zabaras}, {\em Bayesian deep convolutional encoder-decoder
  networks for surrogate modeling and uncertainty quantification}, Journal of
  Computational Physics, 366 (2018), pp.~415--447.

\bibitem{zhu2019physics}
{\sc Y.~Zhu, N.~Zabaras, P.~Koutsourelakis, and P.~Perdikaris}, {\em
  Physics-constrained deep learning for high-dimensional surrogate modeling and
  uncertainty quantification without labeled data}, Journal of Computational
  Physics, 394 (2019), pp.~56--81.

\end{thebibliography}

\end{document}